\documentclass[namedreferences]{SolarPhysics}
\usepackage[optionalrh,solaenum]{spr-sola-addons} 
\usepackage{graphicx}                    
\usepackage{color}                       
\usepackage{url}                         
\usepackage[hyperindex,breaklinks]{hyperref}

\newcommand{\etal}{{\it et al.}}
\newcommand{\insitu}{{\it in-situ}}
\newcommand{\apriori}{{\it a priori}}
\newcommand{\eg}{{\it e.g.}}

\newcommand{\f}[2]{ \frac{#1}{#2} }
\newcommand{\pder}[2]{ \f{\partial #1}{\partial #2} }
\newcommand{\pdder}[2]{ \f{\partial ^{2} #1}{\partial #2 ^{2}} }

\begin{document}

\begin{article}

\begin{opening}

\title{Grad-Shafranov reconstruction of magnetic clouds: overview and improvements}

%
\author{A.~\surname{Isavnin}$^{1}$\sep
        E.~K.~J.~\surname{Kilpua}$^{1}$\sep
        H.~E.~J.~\surname{Koskinen}$^{1, 2}$
       }

%

%
  \institute{$^{1}$ Department of Physics, University of Helsinki, P.O.Box 64, FI-00014, Finland, email: \url{Alexey.Isavnin@helsinki.fi}\\
  			 $^{2}$ Finnish Meteorological Institute, Helsinki, Finland
             }

\begin{abstract}
The Grad-Shafranov reconstruction is a method of estimating the orientation (invariant axis) and cross-section of magnetic flux ropes using the data from a single spacecraft. It can be applied to various magnetic structures such as magnetic clouds (MCs) and flux ropes embedded into the magnetopause and in the solar wind. We develop a number of improvements of this technique and show some examples of the reconstruction procedure of interplanetary coronal mass ejections (ICMEs) observed at 1~AU by the STEREO, WIND and ACE spacecraft during the minimum following the solar cycle 23. The analysis is conducted not only for ideal localized ICME events but also for non-trivial cases of magnetic clouds in fast solar wind. The Grad-Shafranov reconstruction gives reasonable results for the sample events, although it possesses certain limitations, which need to be taken into account during the interpretation of the model results.
\end{abstract}

%

\end{opening}

%

\section{Introduction}\label{s:intro} 

Interplanetary coronal mass ejections (ICMEs) are the heliospheric manifestations of coronal mass ejections (CMEs) at the Sun. Accroding to \inlinecite{Richardson2010} $30\%$ of ICMEs oberved near Earth are MCs. Magnetic clouds as defined by \inlinecite{Burlaga1981} are interplanetary structures with dimensions of the order of $0.25$~AU, which can be identified in {\insitu} spacecraft observations as chunks of solar wind with magnetic field stronger than average, smooth monotonic rotation of the magnetic field through a large angle, low proton temperature and low plasma beta; the most complete up-to-date list of MC signatures can be found in \inlinecite{Zurbuchen2006}. The present concept of MCs assumes a flux rope to be either tied in the Sun forming a magnetic bottle configuration or entirely disconnected from the Sun forming a closed loop. This is proved by observations of bidirectional suprathermal electrons at 1AU \cite{Gosling1987}. The detection of flare associated electrons within MCs supports the magnetic connection to the Sun ({\eg} \inlinecite{Farrugia1993}).

ICMEs are known to cause the strongest magnetospheric disturbances ({\eg} Huttunen {\etal}, \citeyear{Huttunen2002}). MCs can provide strong southward interplanetary magnetic field. To study the geomagnetic efficiency of an MC it is important to know its orientation, shape and size. One can get these critical parameters using various flux rope modelling techniques, first attempts of which were made by \inlinecite{Burlaga1988}. Numerous flux rope models exist today, such as minimum variance analysis (MVA), force-free models such as the Lepping model \cite{Lepping1990}, cylinder and torus models by  \inlinecite{Marubashi2007}, non-force-free eliptical model by \inlinecite{Hidalgo2002}, kinematically distorted model by \inlinecite{Owens2006}, etc. All these models fit {\insitu} observations to an assumed structure of the flux rope cross-section. The Grad-Shafranov reconstruction (GSR) technique, on the contrary, uses spacecraft observations as initial parameters for the reconstruction, thus eliminating the necessity of {\apriori} estimation of the MC boundary. GSR was originally developed for reconstruction of flux ropes embedded into the magnetopause \cite{Hau1999} and later applied to magnetic clouds \cite{Hu2002}. An extented version of GSR useful for multiple spacecraft observations was derived by \inlinecite{Moestl2008}.

In this paper we present improvements to the GSR technique, show examples of its usage and discuss main constraints of the method. In Section \ref{s:gsr} we shortly describe the GSR method and our modifications to it. In Section \ref{s:GSR_examples} we present examples of events reconstructed with the modified GSR technique and in Section \ref{s:Discussion} we discuss and summarize our results.

\section{GSR and improvements}\label{s:gsr} 

The detailed description of the GSR method can be found in \inlinecite{Hau1999} and \inlinecite{Hu2002}. Here we just outline the general algorithm emphasizing its bottlenecks and possible improvements.

GSR uses a number of assumptions. Magnetic clouds passing the observing spacecraft are assumed to be in magnetohydrostatic equilibrium
\begin{equation}  \label{eq:MHSeq}
	\nabla p = \mathbf{j} \times \mathbf{B}
\end{equation}
The magnetic field is assumed to have translation symmetry with respect to an invariant axis direction, i.e. the flux rope is assumed to have $2\frac{1}{2}$-dimensional structure, where the approximation ${\partial}/{\partial z}=0$ can be used. The whole analysis is carried in the deHoffmann-Teller (HT) frame, in which the electric field vanishes and thus the magnetic structure can be treated as time-stationary: ${\partial B}/{\partial t}=0$.

For $2\frac{1}{2}$D magnetic structures with the invariant axis along $z$ equation (\ref{eq:MHSeq}) can be given by the Grad-Shafranov equation
\begin{equation}  \label{eq:GS}
	\pdder{A}{x} + \pdder{A}{y} = - \mu _{0} \f{\mathrm{d}}{\mathrm{d}A} \left( p + \f{B ^{2} _{z}}{2 \mu _{0}} \right), 
\end{equation}
where $A$ is the magnetic vector potential and the magnetic field vector is $\mathbf{B} = \left[ {\partial A}/{\partial y}, -{\partial A}/{\partial x}, B _{z} (A) \right]$. The plasma pressure, the axial magnetic field component and thus the transverse pressure $P_t = p+{B ^{2} _{z}}/{2 \mu _{0}}$ are functions of $A$ alone.

The numerical GS solver is implemented using the Taylor expansions: 
\begin{equation}  \label{eq:GS_solver_A}
	A(x, y \pm \Delta y) \simeq A(x, y) + \left( \pder{A}{y} \right) _{x, y}(\pm \Delta y) + \f{1}{2} \left( \pdder{A}{y} \right) _{x, y} (\pm \Delta y) ^{2}
\end{equation}
\begin{equation}  \label{eq:GS_solver_Bx}
	B _{x} (x, y \pm \Delta y) \simeq B _{x}(x, y) + \left( \pdder{A}{y} \right) _{x, y} (\pm \Delta y)
\end{equation}
Equation (\ref{eq:GS}) expresses an implicit Cauchy problem  with a numerical solver formed by equations (\ref{eq:GS_solver_A}, \ref{eq:GS_solver_Bx}). So what we have here is an ill-defined problem without boundary conditions, which is subject to growth of singularities after certain number of steps of this recurrent algorithm.

Originally in \inlinecite{Hau1999} a finite difference approximation was used to estimate ${\partial^2 A}/{\partial x^2}$, which is very unstable when dealing with noisy data, as most experimental data are. 

\begin{figure}
\centerline{\includegraphics[width=0.8\textwidth,clip=]{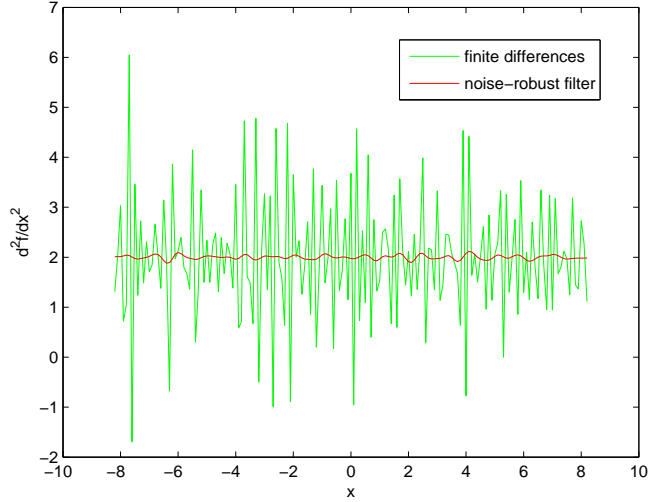}}
\caption{Finite differences method vs noise-robust filter for estimation of the second order derivative.}\label{fig:fd_VS_nrf}
\end{figure}

In this paper we use smooth noise-robust differentiators \cite{Holoborodko2008} to improve the stability of the algorithm. Standard finite difference schemes, such as central differences, lack high-frequency suppression and may result in erroneous results when estimating derivatives in case of noisy data. Noise-robust filters, on the contrary, guarantee suppression of high-frequency noise. The existing numerical scheme of solving the GS equation (\ref{eq:GS}) suggests multiple differentiation of initial data, so that instabilities caused by numerical differentiation grow like an avalanche. Therefore the use of noise-robust filter can suppress, at least to some extent, the growth of such singularities. In Figure~\ref{fig:fd_VS_nrf} the numerical second derivative of the function $f(x)=x^2+NL*RANDN$, where $NL$ is the noise level and $RANDN$ is a random pick from the normal distribution, is shown. From such a simple example it is clear that a noise-robust differentiator of the second order gives a much more stable result.

The determination of the invariant axis is a critical point in the whole reconstruction procedure. It is based on the assumption of constant transverse pressure and constant magnetic vector potential on common magnetic field lines. The search for invariant axis is performed by trial and error. For each test direction of the axis magnetic field data are projected to the plane perpendicular to the axis. The transverse pressure and magnetic potential are calculated in this plane. For the best-fit direction of the invariant axis of the flux rope the $P_t(A)$ curve forms two coinciding branches. These two branches represent the motion of the spacecraft inward and outward of the flux rope, which correspond to the decrease and increase of the distance between the spacecraft and the invariant axis, respectively. Obviously, the point of the $P_t(A)$ curve, which connects two branches, correspond to the closest approach of the spacecraft to the invariant axis. Results of this search are visualized as a residual map. For each test direction the residue between inward and outward branches is calculated using the equation
\begin{equation} \label{eq:residue}
	\mathcal{R} = \left[ \displaystyle \sum \limits_{i=1}^{m_0} \left( P _{t, i} ^{\mathsf{in}} - P _{t, i} ^{\mathsf{out}} \right) ^{2} \right] ^{\f{1}{2}} / | \mathsf{max}(P _t) - \mathsf{min}(P _t) |
\end{equation}

Essentially, the residual map is a contour plot of the residue on top of a hemisphere, that represents all possible directions of the invariant axis. The initial coordinates used in the residual map are usually defined as follows: $\mathbf{\hat{y}}$ is the direction of the maximum variance of the magnetic field, constrained by \mbox{$\mathbf{\hat{y}} \cdot \mathbf{V_{HT}}=0$}, \mbox{$\mathbf{\hat{x}} = -\mathbf{\hat{V}_{HT}}$}, \mbox{$\mathbf{\hat{z}}=\mathbf{\hat{x}}\times\mathbf{\hat{y}}$}. The direction at $0^{\circ}$ longitude and $90^{\circ}$ latitude is $\mathbf{\hat{x}}$, the direction at $90^{\circ}$ longitude and $90^{\circ}$ latitude is $\mathbf{\hat{y}}$ and the direction at $0^{\circ}$ latitude is $\mathbf{\hat{z}}$. The search for the invariant axis is performed by stepping away from $\mathbf{\hat{z}}$ and calculating residue for each trial direction. We obtain the trial direction of the invariant axis in original coordinate system (i.e. GSE, RTN, etc.) on each step by rotating $\mathbf{\hat{z}}$ by corresponding latitude and longitude angles. 

\begin{figure}
\centerline{
\includegraphics[width=0.5\textwidth,clip=]{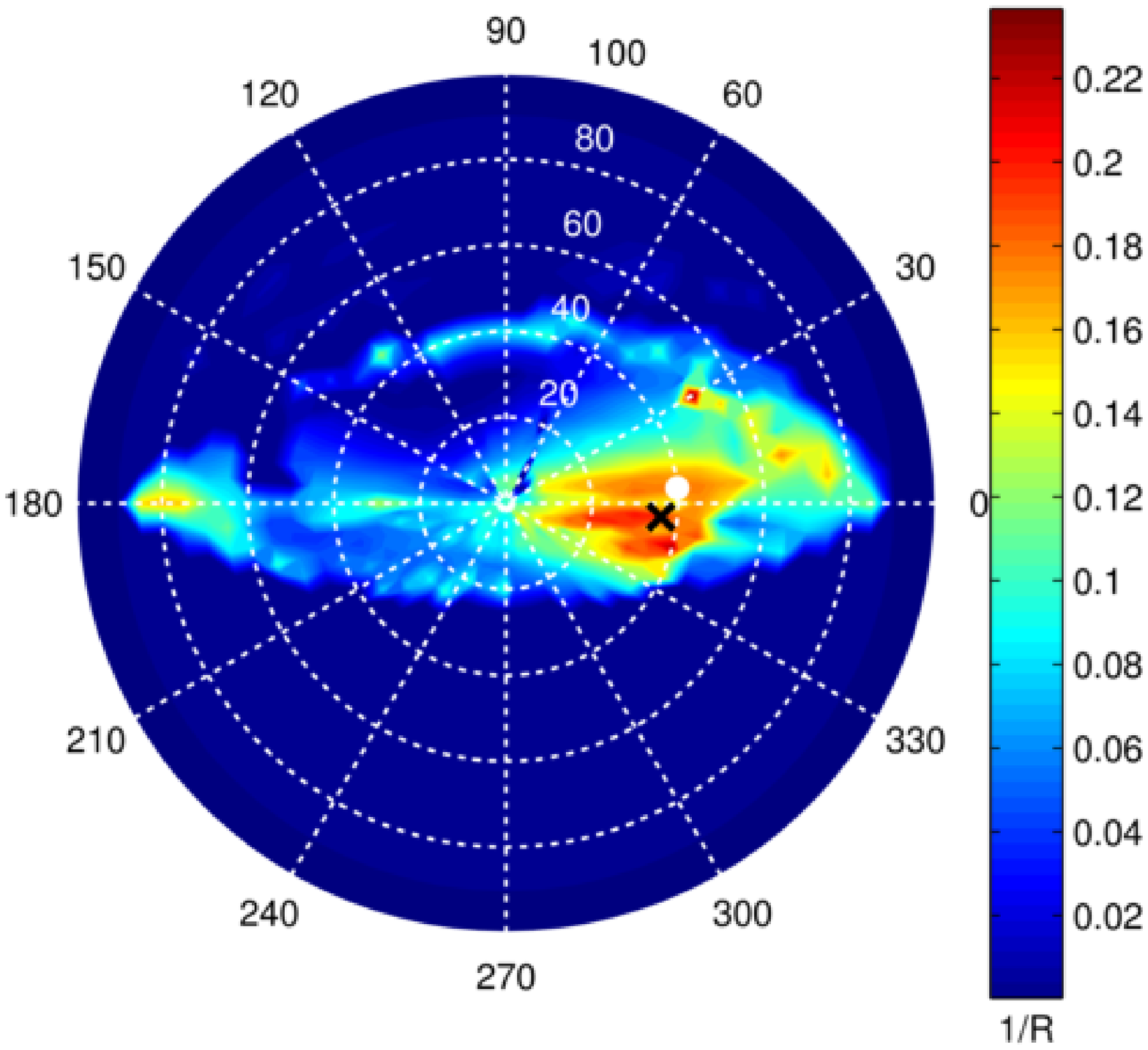}
\includegraphics[width=0.5\textwidth,clip=]{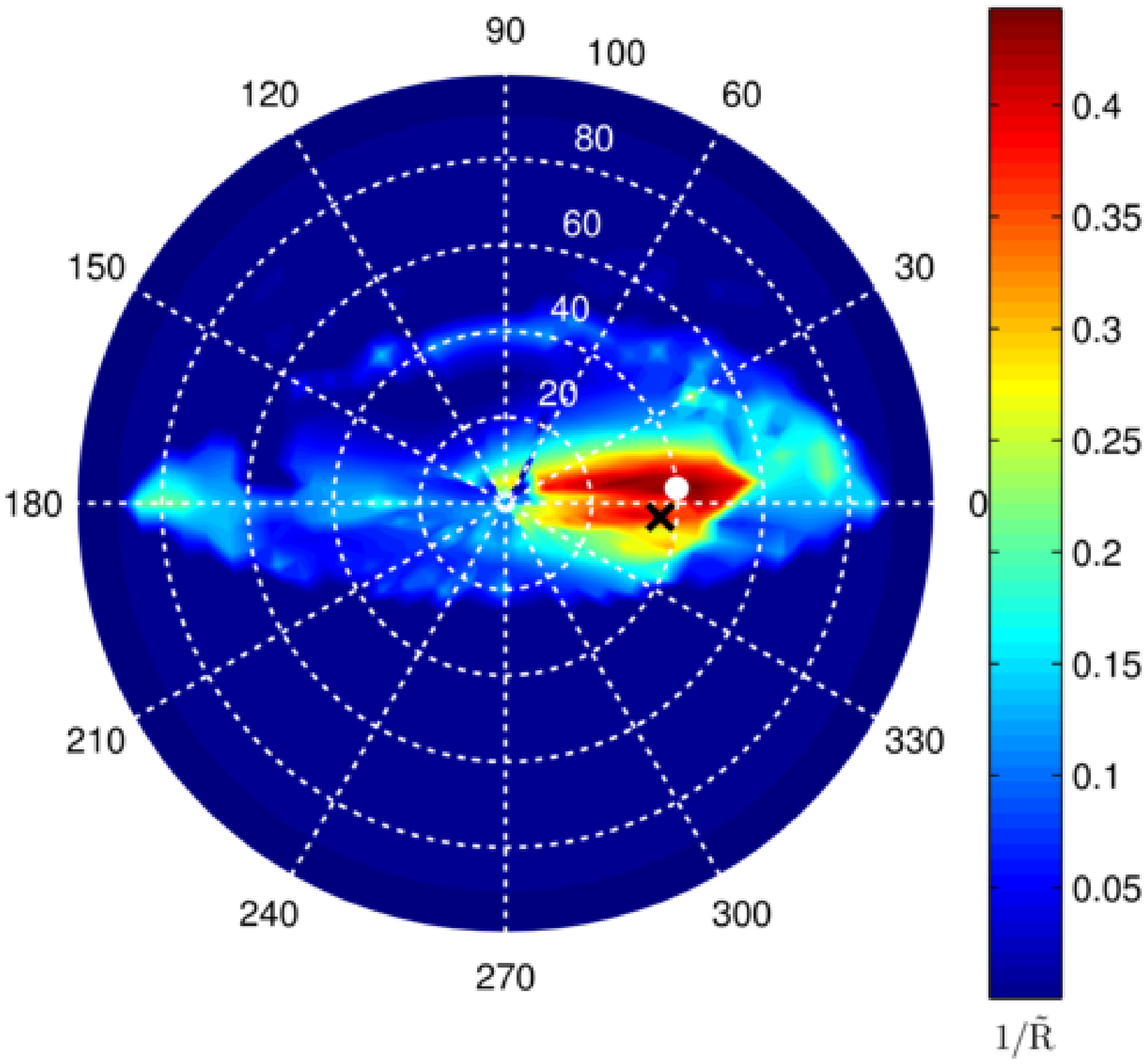}
}
\caption{Residual map for the 1995-10-18 WIND event: original (left) and filtered (right). White dot marks the GSR-optimized direction of the invariant axis. Black cross marks the direction of intermediate variance of the magnetic field vector.}\label{fig:WIND_1995-10-18_RM}
\end{figure}

An example of the residual map is shown in Figure~\ref{fig:WIND_1995-10-18_RM}. The black cross shows the direction of intermidiate variance of magnetic field vector. The direction with the minimum residue is considered to be the invariant axis direction and is denoted by the thick white dot on the residual map. A problem of this method is that the residual maps are occasionally saturated with false possible axis directions that correspond to short lengths of branches of the $P_t(A)$ curve and thus smaller values of the residue. An example of this is given in Figure~\ref{fig:WIND_1995-10-18_RM} (left).

To eliminate this issue we combine the residue map with a branch length map. The final combined residue is calculated as
\begin{equation} \label{eq:RL}
	\tilde{\mathcal{R}} = \mathcal{R} \f{N}{2L}, 
\end{equation}
where L is the length of the branches (in terms of number of data points) and N is the number of observational data points. So ${2L}/{N}$ in (\ref{eq:RL}) is a fraction of the initial data interval where the coincidence of the branches of $P_t(A)$ takes place. Obviously, for the optimal direction of the invariant axis ${2L}/{N}$ shows the fraction of the initial data interval occupied by the flux rope. The comparison of the original residual map and filtered by equation (\ref{eq:RL}) is shown on Figure~\ref{fig:WIND_1995-10-18_RM} using a WIND event observed on 1995-10-18 as an example. We are using the inverse residue ${1}/{\tilde{\mathcal{R}}}$ in the residue maps. Several false minimum residue directions are present in Figure~\ref{fig:WIND_1995-10-18_RM} (left) (multiple red zones). After filtering with (\ref{eq:RL}) we end up with a much more clear combined residue map (Figure~\ref{fig:WIND_1995-10-18_RM} (right)) with the minimum residue direction approximately in the center of the red zone.

\section{Examples of reconstructed events}\label{s:GSR_examples}

In this section we analyze several events using the GSR technique.

\inlinecite{Jian2006} presented the classification of MCs based on total perpendicular pressure profile. The total perpendicular pressure is defined as the sum of the total magnetic pressure (since the magnetic field does not generate pressure force parallel to the mangetic field direction) and the thermal pressure perpendicular to the magnetic field direction:
\begin{equation} \label{eq:Ptot}
	P_{\bot} = \f{B^2}{2 \mu _0} + \sum \limits_{j} n_j k T_{j_{\bot}},
\end{equation}
where $j$ represents different sorts of particles. MCs were divided in three groups depending on the shape of the pressure profile: Group 1 with a well-determined peak, Group 2 with a plateau observed and Group 3 characterised by decreasing pressure. The three groups correspond to a small, medium and large impact parameters (closest approach of the spacecraft to the axis of the flux rope) respectively. We are checking this property throughout our sample events.

For STEREO events we use the RTN coordinates. In the RTN coordinates, R points from the center of the Sun through the spacecraft. T is formed by the cross product of the solar rotation axis and R, and lies in the solar equatorial plane. N is formed by the cross product of R and T and is the projection of the solar rotational axis on the plane of the sky.

\subsection{STEREO-A event on 2008-11-07}\label{ss:STA_2008-11-07}

\begin{figure}
\centerline{
\includegraphics[width=0.8\textwidth,clip=]{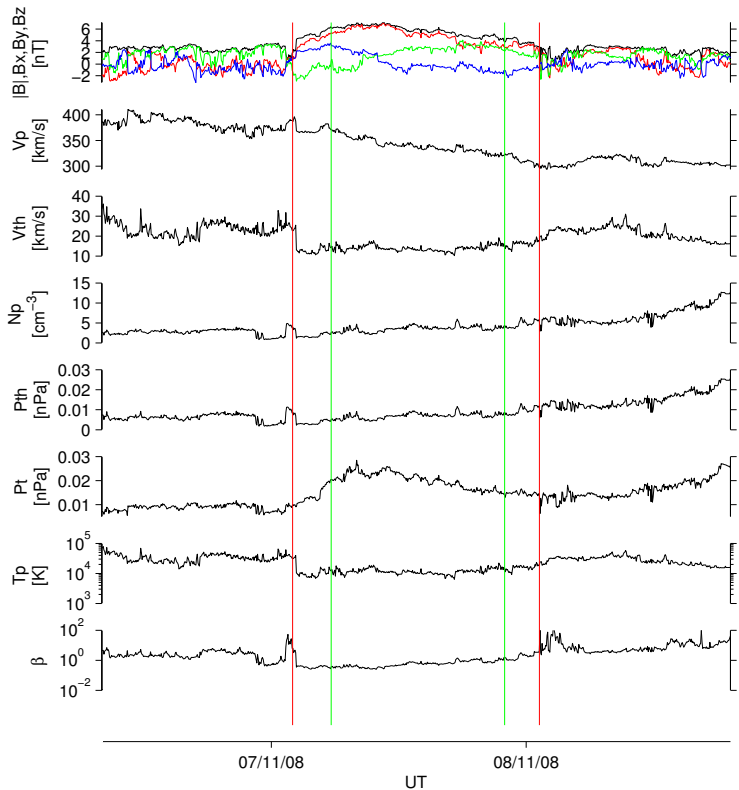}
}
\caption{Magnetic field and plasma data for the 2008-11-07 STA event. Red vertical lines show initial time limits of the MC, green vertical lines show limits of the flux rope as seen in the GS reconstructed magnetic field map. The panels show (from top to bottom): magnetic field magnitude (black) and magnetic field components in the RTN coordinates (red: $B_r$, green: $B_t$, blue: $B_n$), plasma bulk flow speed, proton thermal speed, proton density, plasma thermal pressure, total perpendicular pressure of plasma, proton temperature and plasma beta.}\label{fig:STA_2008-11-07_data}
\end{figure}

\begin{figure}
\centerline{
\includegraphics[width=0.5\textwidth,clip=]{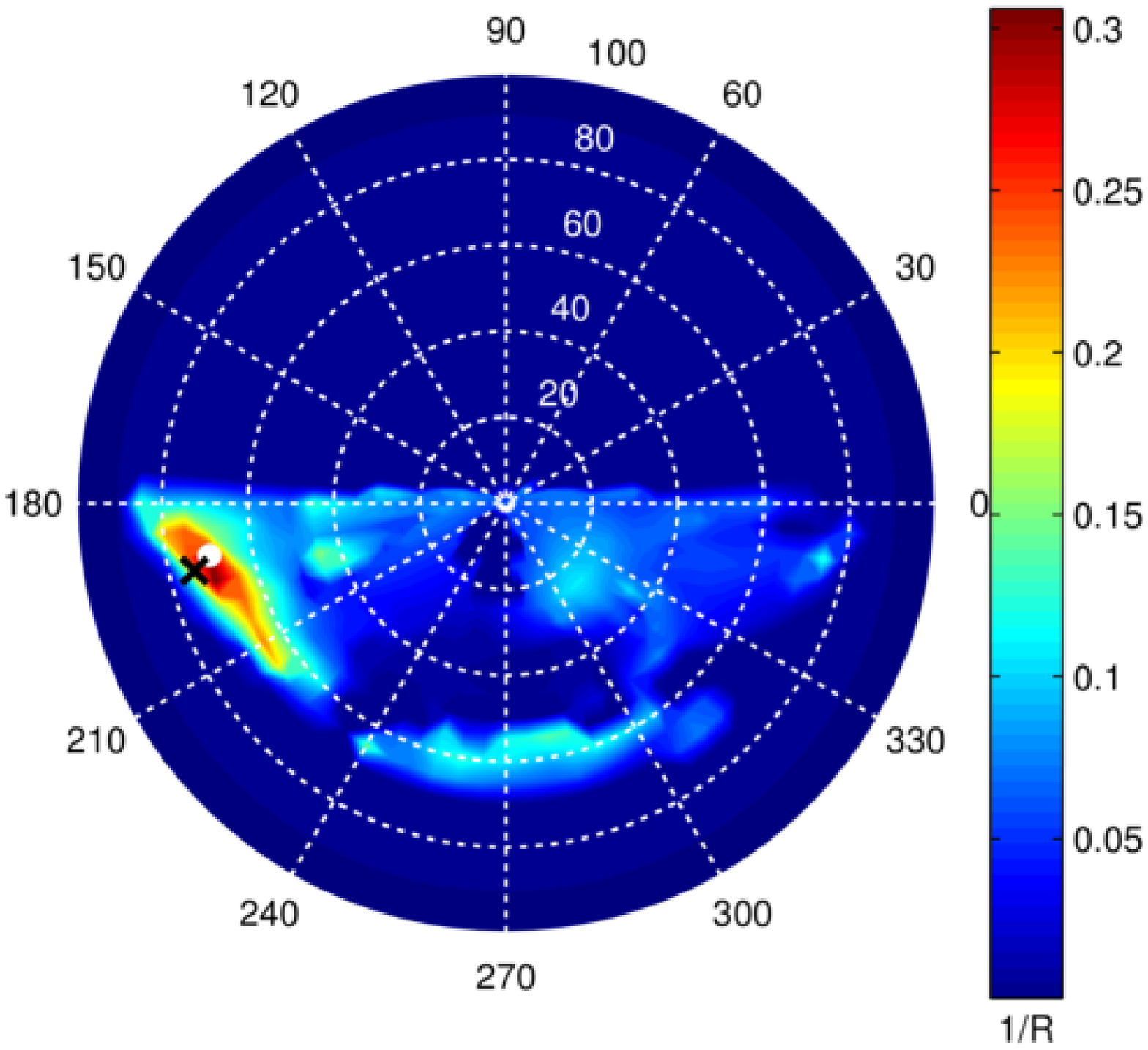}
\includegraphics[width=0.5\textwidth,clip=]{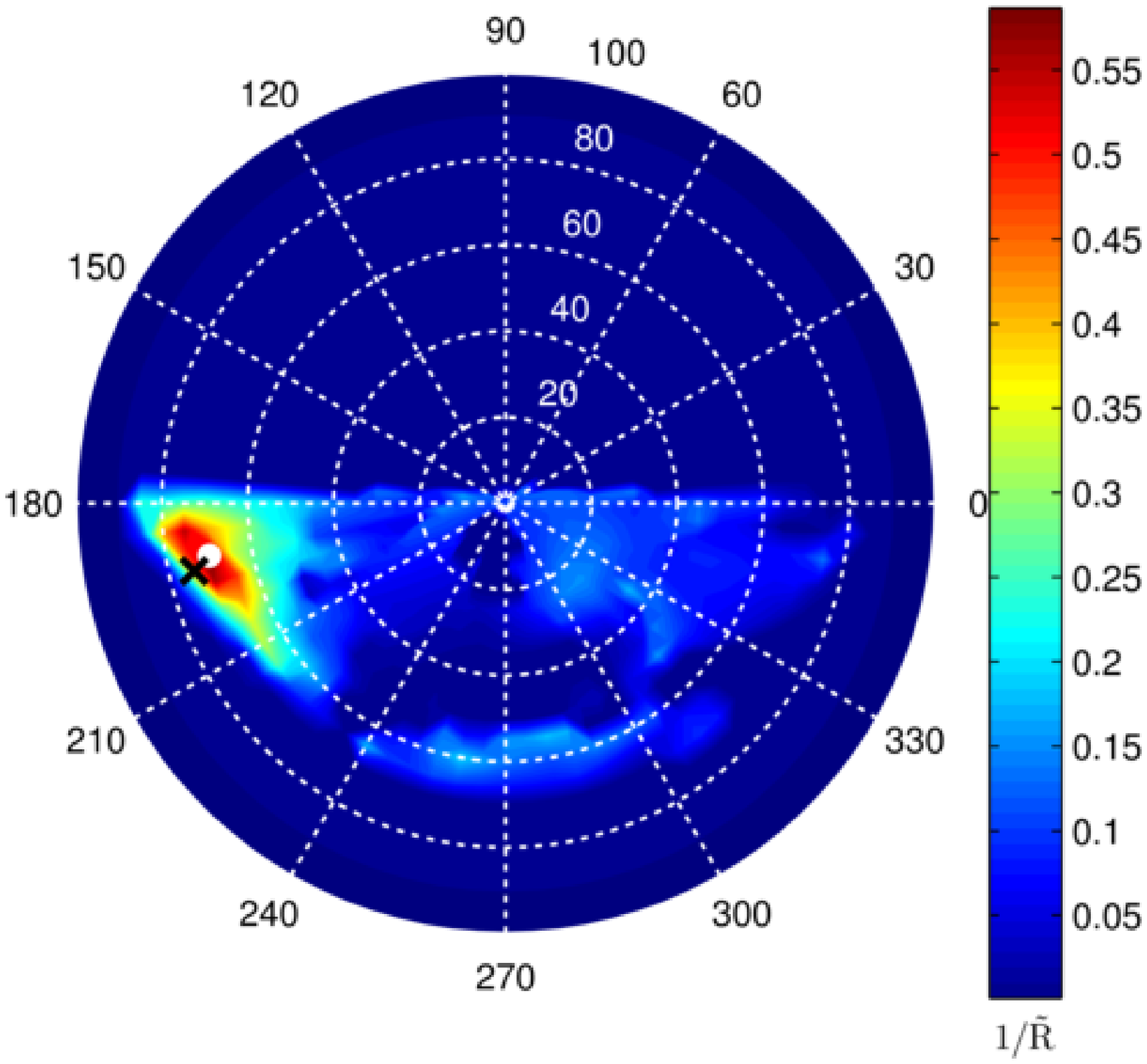}
}
\caption{Original (left) and filtered (right) residue maps for the 2008-11-07 STA event. White dot marks the GSR-optimized direction of the invariant axis. Black cross marks the direction of intermidiate variance of magnetic field vector.}\label{fig:STA_2008-11-07_RM}
\end{figure}

\begin{figure}
\centerline{
\includegraphics[width=0.5\textwidth,clip=]{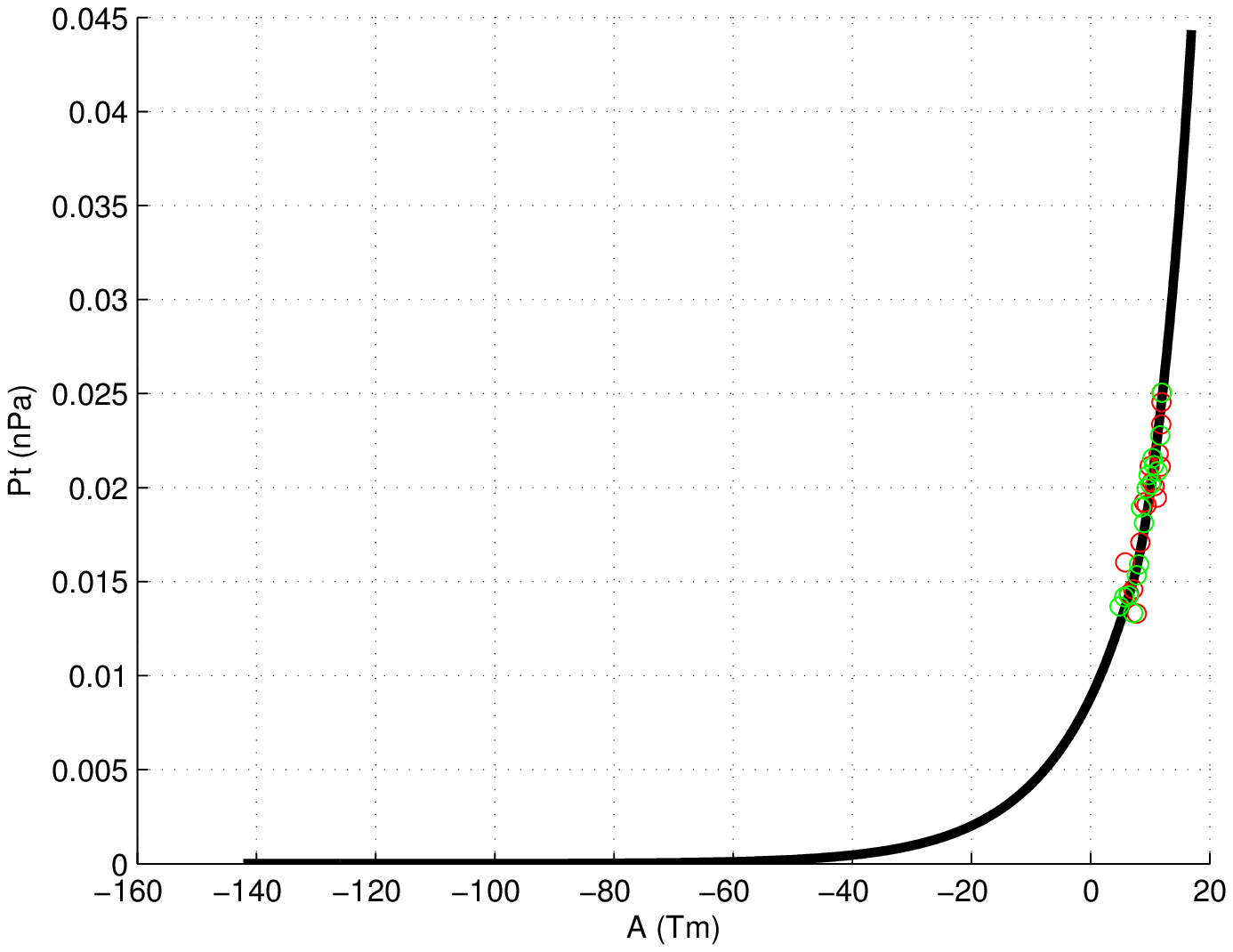}
\includegraphics[width=0.5\textwidth,clip=]{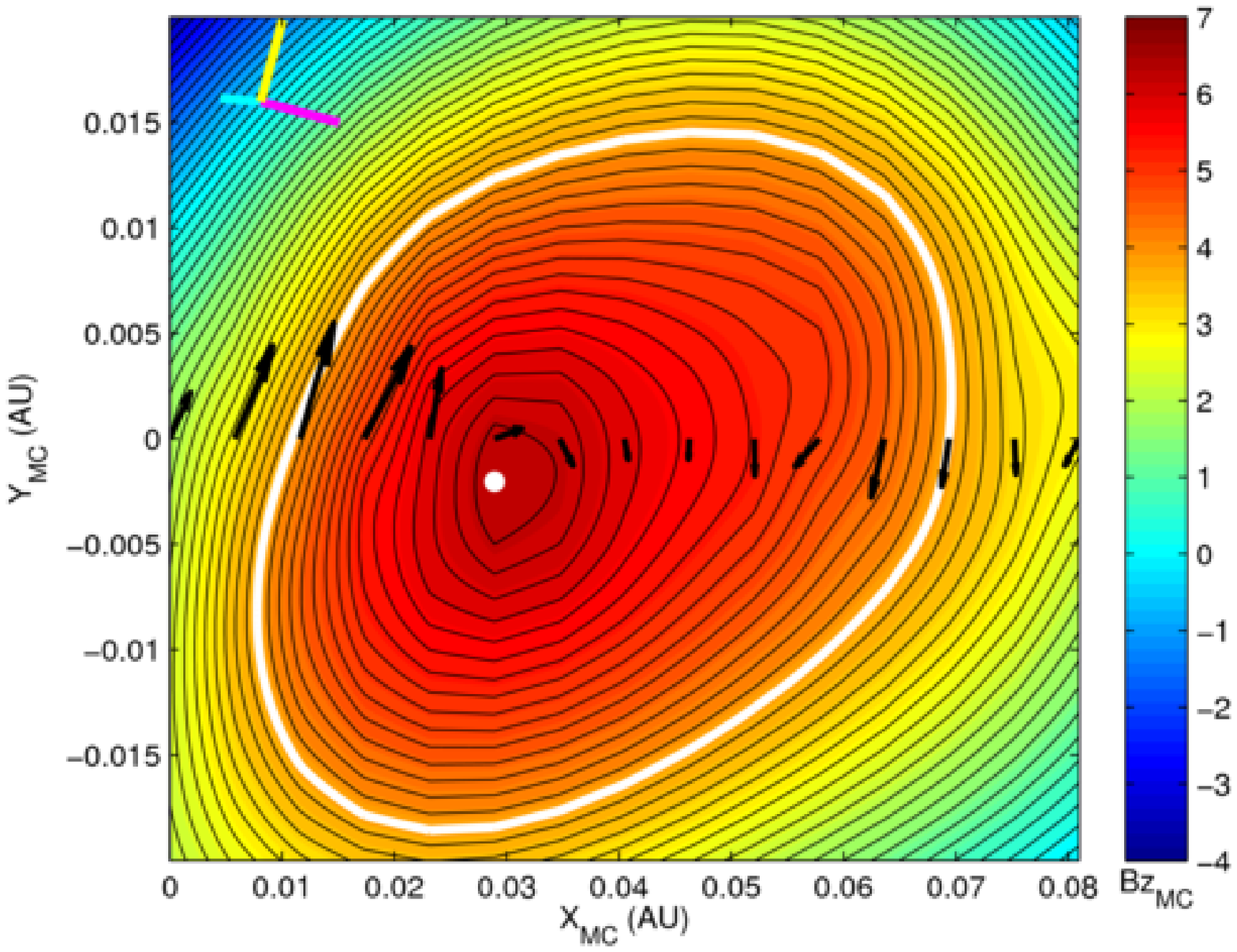}
}
\caption{ $P_t(A)$ curve fitted with polynomial of 3rd order and exponential tail (left) and the reconstructed magnetic field map for the 2008-11-07 STA event. The projected coordinates shown in the magnetic field map are $R$ (cyan), $T$ (magenta), $N$ (yellow).}\label{fig:STA_2008-11-07_AP_GS}
\end{figure}

The first event we analyze is a relatively well-defined MC. It was observed on November 7th 2008 by the STEREO-A spacecraft (Figure~\ref{fig:STA_2008-11-07_data}). This MC shows typical signatures in {\it in situ} data: smooth (i.e. low variance) magnetic field rotation, declining velocity profile (caused by expansion), low proton temperature and low plasma beta. 

The MC originated from a CME event that took place on November 2nd 2008 at 04:00. The deHoffmann-Teller frame speed estimated for this ICME is $V_{HT}=[348.0; 3.0; -13.6]$~km/s in the RTN coordinates with the correlation coefficient $c=0.998$ \cite{Khrabrov1998}. According to the residual map (Figure~\ref{fig:STA_2008-11-07_RM}) the invariant direction of the flux rope is $\theta_{RTN}=4.3^{\circ}$ and $\varphi_{RTN}=26.1^{\circ}$ in the RTN coordinates. The apex of the MC is the part of the flux rope that is furthest from the Sun at any moment of the propagation through the interplanetary space. It is natural to expect that the estimated invariant axis of the flux rope would tend to be almost perpendicular to the radial flow of the solar wind when the spacecraft crosses the flux rope close to its apex. And in turn, when the spacecraft crosses the flux rope far from its apex, i.e. penetrates through one of its legs, the estimated invariant axis tends to be parallel to the radial outflow from the Sun. For this particular event the RTN longitude of the invariant axis signifies that the spacecraft intersects the MC far from its apex. The spacecraft crossed this flux rope close to its axis with the impact parameter of $0.005$~AU. This also agrees with the analysis of total prependicular pressure profile (Figure~\ref{fig:STA_2008-11-07_data}), according to which this event falls into Group 1 in the \inlinecite{Jian2006} classification. The corresponding $P_t(A)$ fitting curve for this direction is plotted in Figure~\ref{fig:STA_2008-11-07_AP_GS} (left). Red circles show the {\it in situ} $P_t(A)$ data when the spacecraft moved inward the flux rope, green circles when it moved outward.

The magnetic field map for this event is shown in Figure~\ref{fig:STA_2008-11-07_AP_GS} (right). It is plotted in the MC coordinate system. The picture is the cross-section of the flux rope in the plane perpendicular to the invariant axis. The $x$-axis is determined by the spacecraft trajectory projected to this plane, black arrows show spacecraft {\it in situ} observations of the magnetic field projected on the same plane. The Sun is to the right of the picture. In the upper left corner of the map the projected RTN coordinate system is shown. Solid black lines are magnetic equipotential lines. The thick white dot, the central point of the flux rope, is the point of the maximum magnetic potential. The thick white line is the boundary of the flux rope, defined as the absolute minimum of the magnetic potential for which two branches of the $P_t(A)$ curve still coincide. The area constrained by this boundary may be thought of as an area of reliable GS reconstruction, since for the outer part of the magnetic field map the fitting curve for $P_t(A)$ is extrapolated. Using this boundary it is possible to get the time limits for the spacecraft passage through the flux rope. These time limits are marked with green vertical lines in Figure~\ref{fig:STA_2008-11-07_data} and are narrower than those that were estimated by visual analysis. This is actually an output parameter of the GSR technique. The first estimate of the time limits of the reconstructed event by visual analysis of magnetic field and plasma data may be rough, since, unlike the flux rope fitting models, this method is not that sensitive to the choice of the time boundaries for the MC measurements. Note, that for such a well-defined MC the minimum variance analysis (MVA) gives the direction of the invariant axis very close to the direction estimated with GSR.

\subsection{ACE/WIND event on 2004-11-09: fast ICME}\label{ss:ACE_WIND_2004-11-09}

\begin{figure}
\centerline{
\includegraphics[width=0.8\textwidth,clip=]{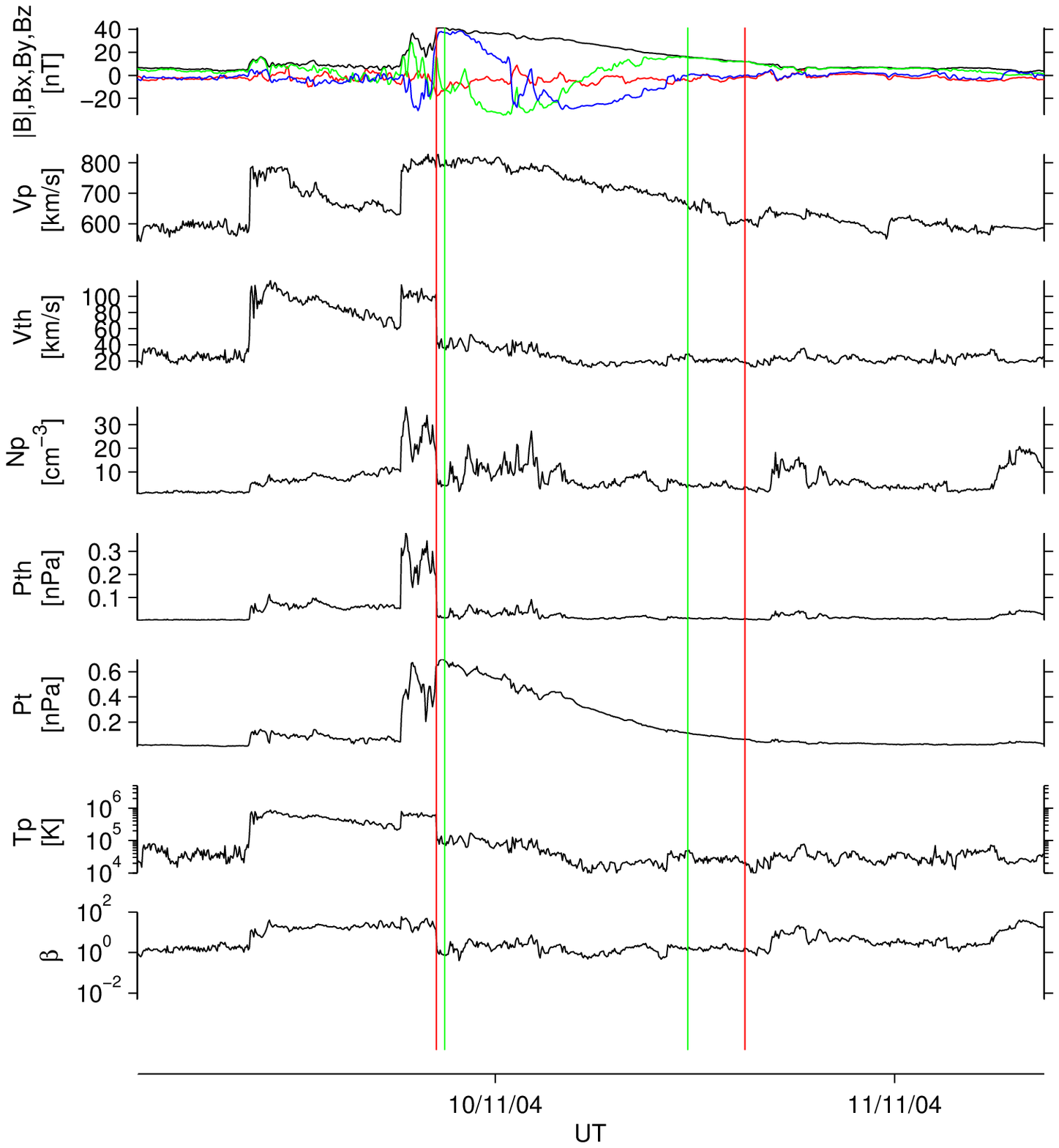}
}
\centerline{
\includegraphics[width=0.8\textwidth,clip=]{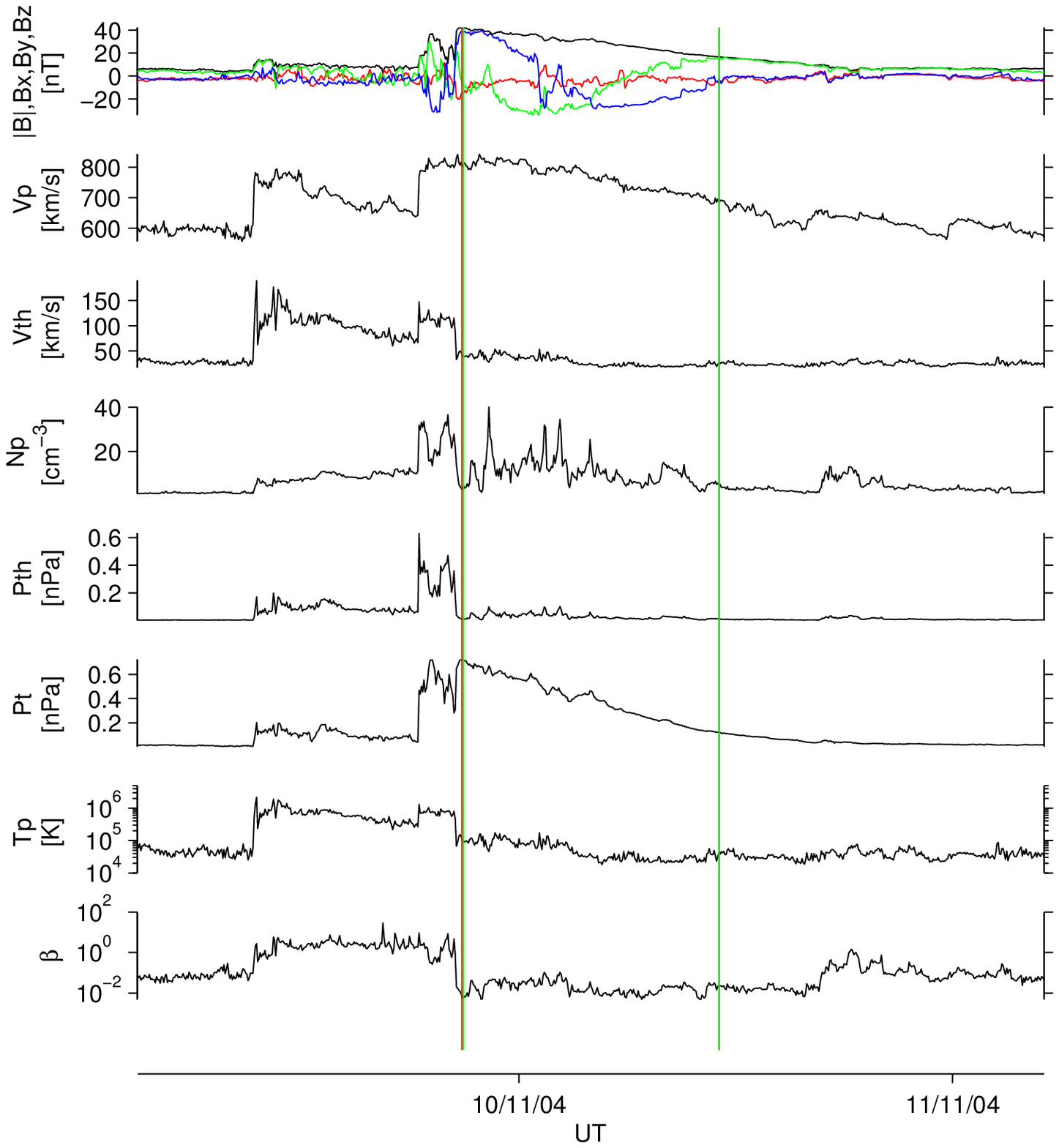}
}
\caption{Magnetic field and plasma data for the 2004-11-09 ACE(top)/WIND(bottom) event. Red vertical lines show initial time limits of MC, green vertical lines show limits of the flux rope as seen in the GS reconstructed magnetic field map. The notations are the same as in Figure~\ref{fig:STA_2008-11-07_data}.}\label{fig:ACE_WIND_2004-11-09_data}
\end{figure}

\begin{figure}
\centerline{
\includegraphics[width=0.5\textwidth,clip=]{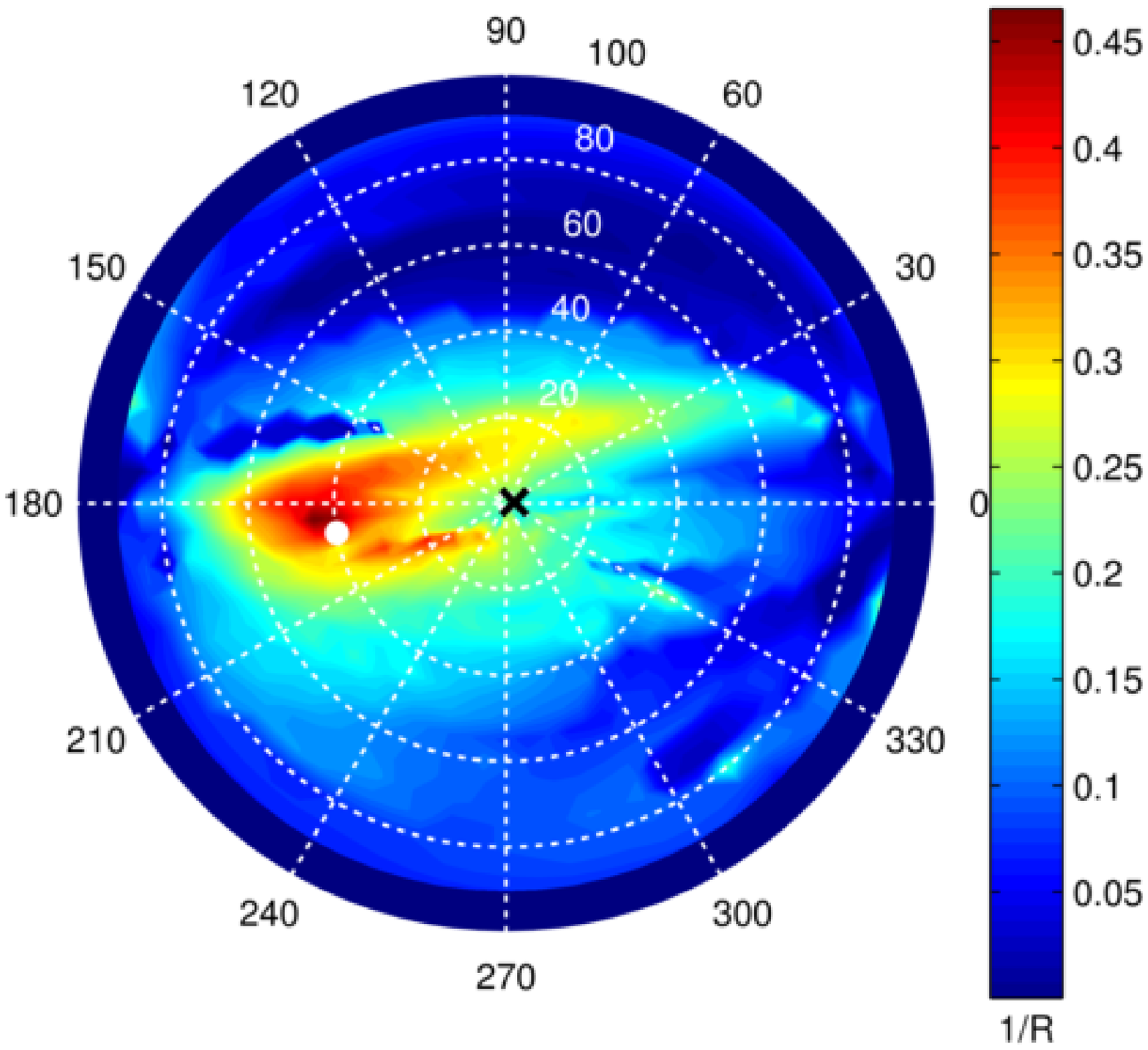}
\includegraphics[width=0.5\textwidth,clip=]{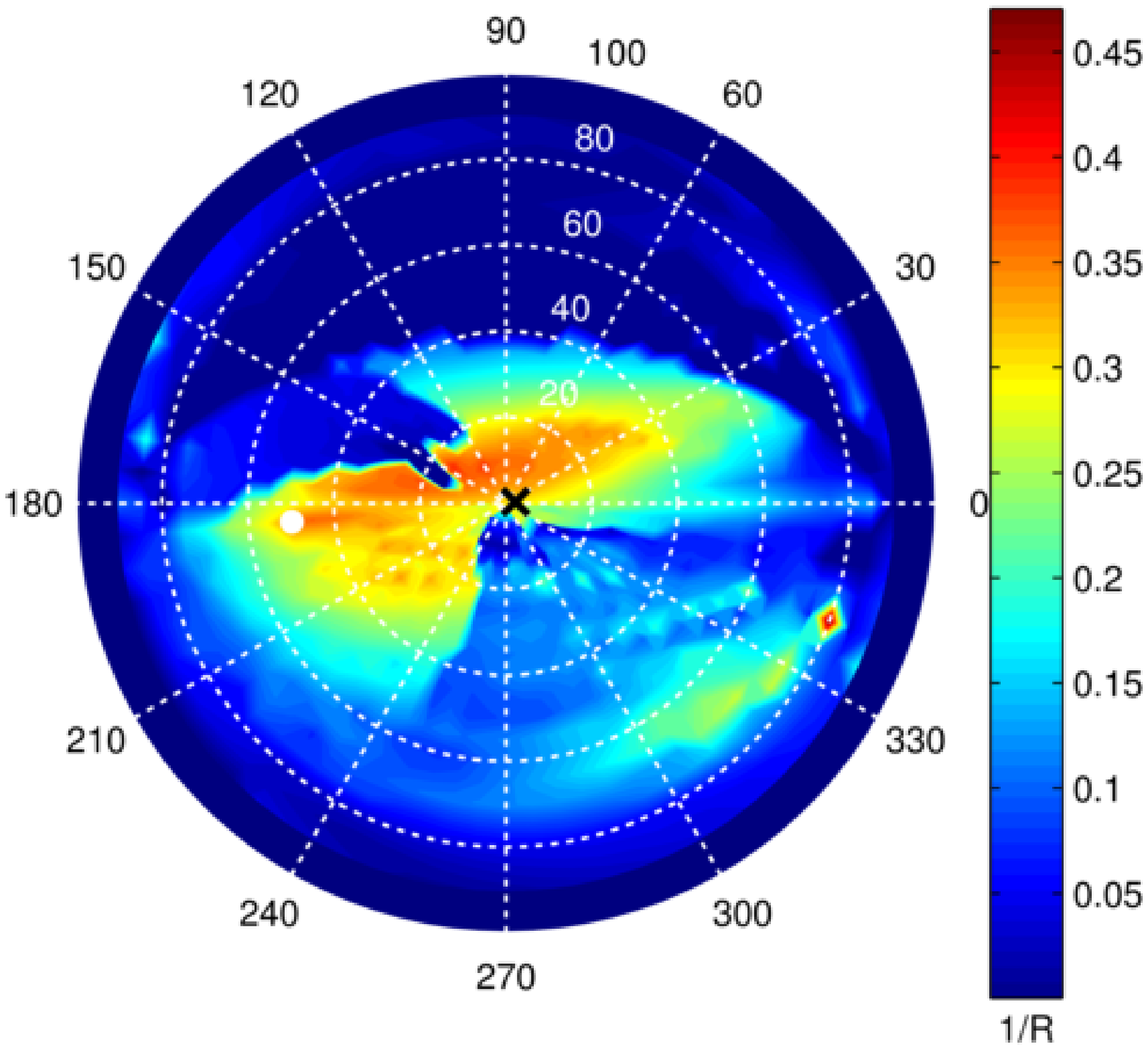}
}
\centerline{
\includegraphics[width=0.5\textwidth,clip=]{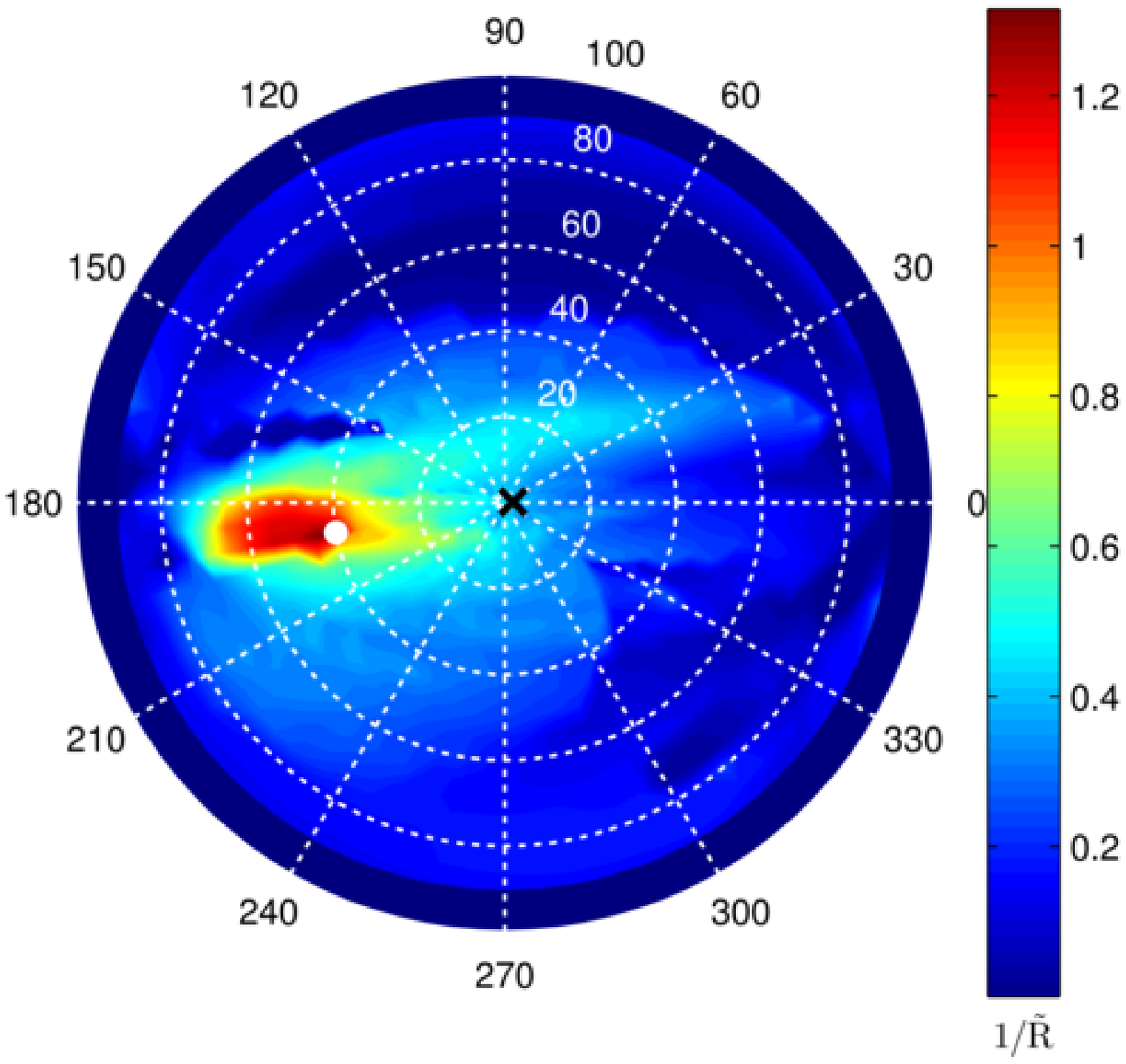}
\includegraphics[width=0.5\textwidth,clip=]{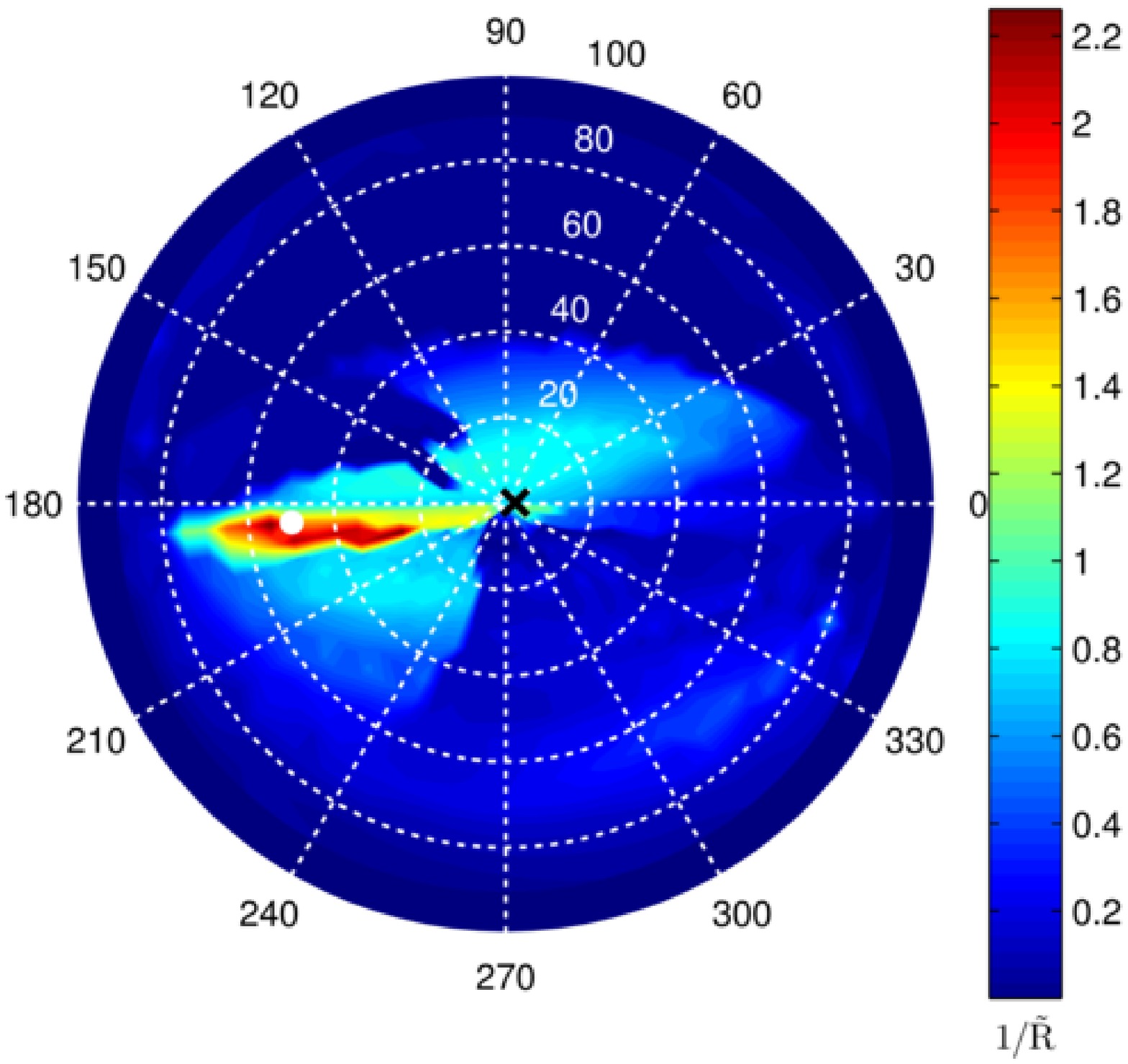}
}
\centerline{
\includegraphics[width=0.5\textwidth,clip=]{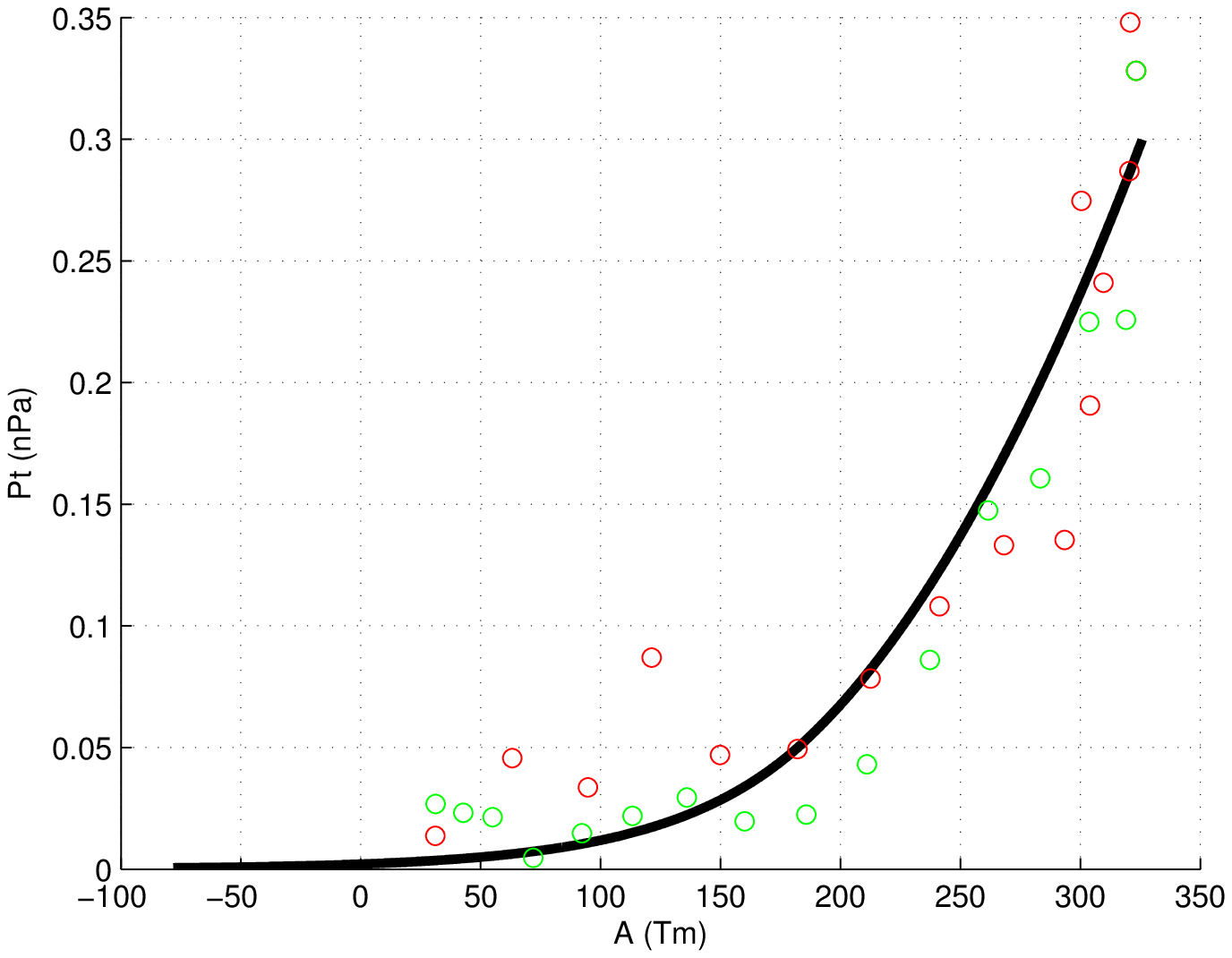}
\includegraphics[width=0.5\textwidth,clip=]{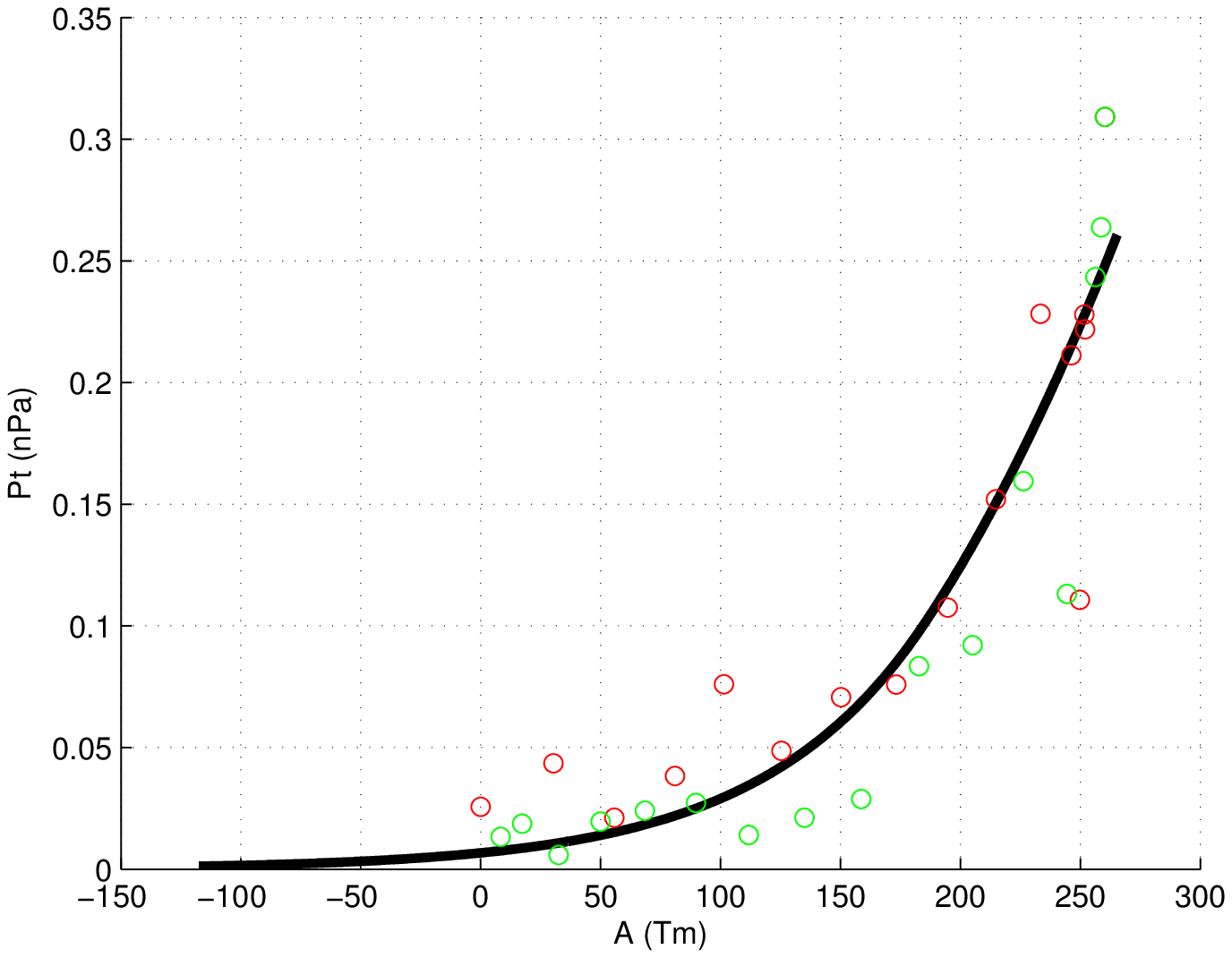}
}
\caption{Original (top) and filtered (middle) residual maps and $P_t(A)$ curves (bottom) for ACE (left) and WIND (right) for the 2004-11-09 event.}\label{fig:ACE_WIND_2004-11-09_RM_AP}
\end{figure}

\begin{figure}
\centerline{
\includegraphics[width=0.5\textwidth,clip=]{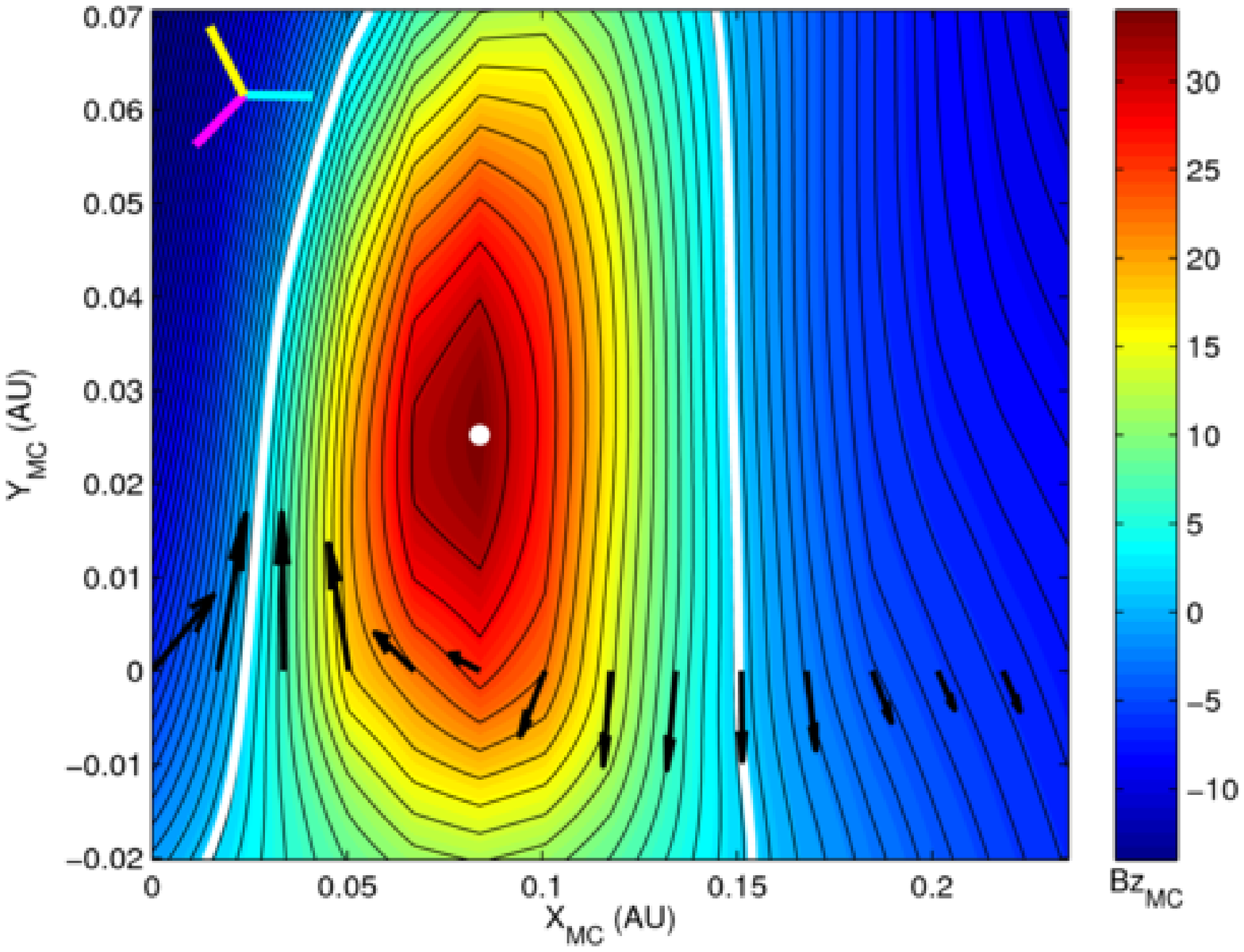}
\includegraphics[width=0.5\textwidth,clip=]{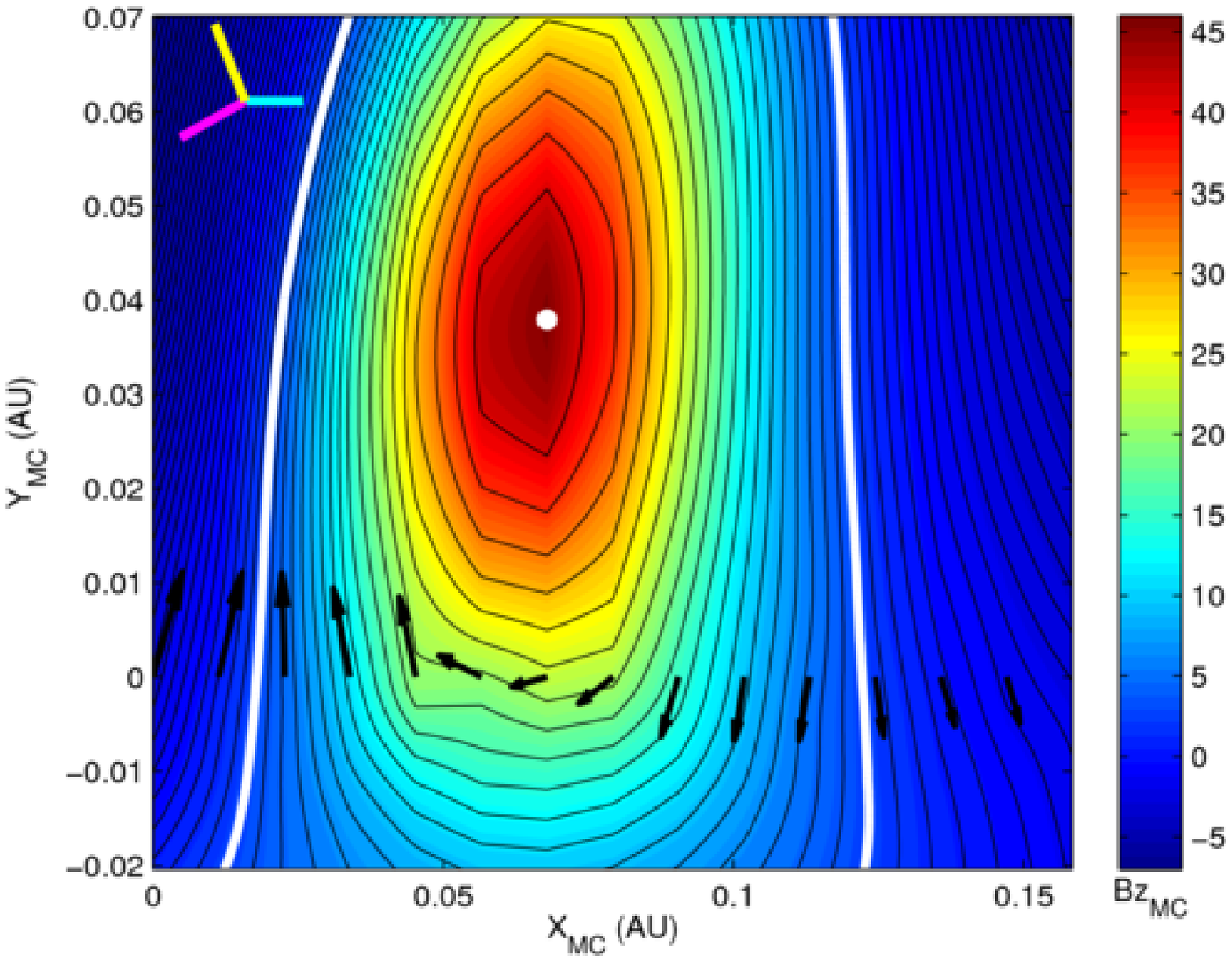}
}
\caption{Reconstructed magnetic field map for ACE (left) and WIND (right) for the 2004-11-09 event. The projected coordinates shown in the magnetic field maps are $X_{GSE}$ (cyan), $Y_{GSE}$ (magenta), $Z_{GSE}$ (yellow).}\label{fig:ACE_WIND_2004-11-09_GS}
\end{figure}

The MC observed on November 9th 2004 at L1 is an example of a fast ICME with the average speed higher than $750$~km/s (Figure~\ref{fig:ACE_WIND_2004-11-09_data}). This MC originated from a CME observed on 2004-11-07 at 16:55 UT. Since the propagation speed of this ICME well exceeded the average speed of the solar wind, it would be expected to produce a leading shock. The analyzed MC was, however, associated with two leading shocks, detected by WIND at 09:19 UT and 18:25 UT on November 9. Here we reconstruct this MC using data samples obtained by the ACE and WIND spacecraft. While multispacecraft reconstruction was already performed by \inlinecite{Moestl2008}, we aim here on comparison of these two separate reconstruction procedures and their stability.

Using the residual map analysis (Figure \ref{fig:ACE_WIND_2004-11-09_RM_AP}) we estimate the invariant axis direction to be \mbox{$\theta_{GSE}=-27.0^{\circ}$}, \mbox{$\varphi_{GSE}=42.9^{\circ}$} for ACE and \mbox{$\theta_{GSE}=-15.8^{\circ}$}, \mbox{$\varphi_{GSE}=36.5^{\circ}$} for WIND. This MC was crossed by the spacraft rather far from its apex, but closer than the MC analyzed in previous section.

According to \inlinecite{Jian2006} the MCs characterised by the decrease of total perpendicular pressure in spacecraft data (as seen in Figure~\ref{fig:ACE_WIND_2004-11-09_data}) belong to Group 3 type of events with the flux rope axis relatively far from the spacecraft trajectory, which is proved by our reconstruction results (Figure~\ref{fig:ACE_WIND_2004-11-09_GS}). The impact parameter for this magnetic cloud is $0.025$~AU for ACE and $0.037$~AU for WIND. The distance between the spacecraft as seen from magnetic field maps is about $0.008$~AU which is of the order of the mean distance in the $YZ$ plane in GSE between ACE and WIND, which in turn is equal to $0.002$~AU during the event. The step of reconstruction along the $y$-axis used for this event is $\Delta y = 0.0001$~AU, which means that the lack of resolution cannot be the reason of this difference. The possible causes for this difference are the inequality between the invariant axis directions obtained for two spacecraft and the inaccuracy of the GSR method.

According to Figure~\ref{fig:ACE_WIND_2004-11-09_GS} this MC embeds only one flux rope though two leading shocks were observed. Note also, that the invariant axis directions estimated by MVA and GSR techniques differ by $\sim 40-50^{\circ}$ for this event.

\subsection{STEREO-A event on 2009-07-11: ICME followed by a stream interaction region}\label{ss:STA_2009-07-11}

\begin{figure}
\centerline{
\includegraphics[width=0.8\textwidth,clip=]{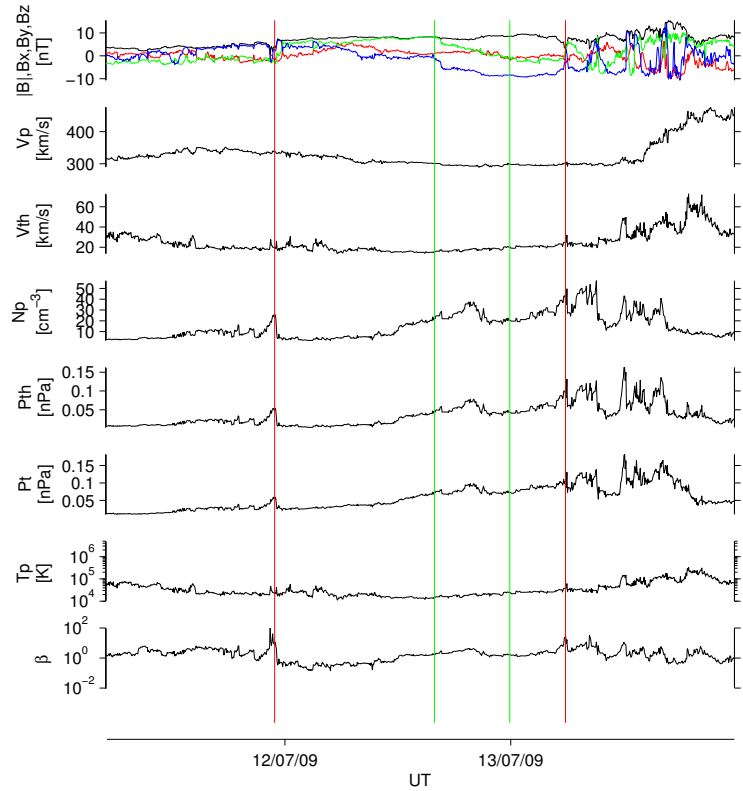}
}
\caption{Magnetic field and plasma data measured on STEREO-A for the 2009-07-11 event. Red vertical lines show initial time limits of the MC, green vertical lines show limits of the flux rope as seen in the GS reconstructed magnetic field map. The notations are the same as in Figure~\ref{fig:STA_2008-11-07_data}.}\label{fig:STA_2009-07-11_data}
\end{figure}

\begin{figure}
\centerline{
\includegraphics[width=0.5\textwidth,clip=]{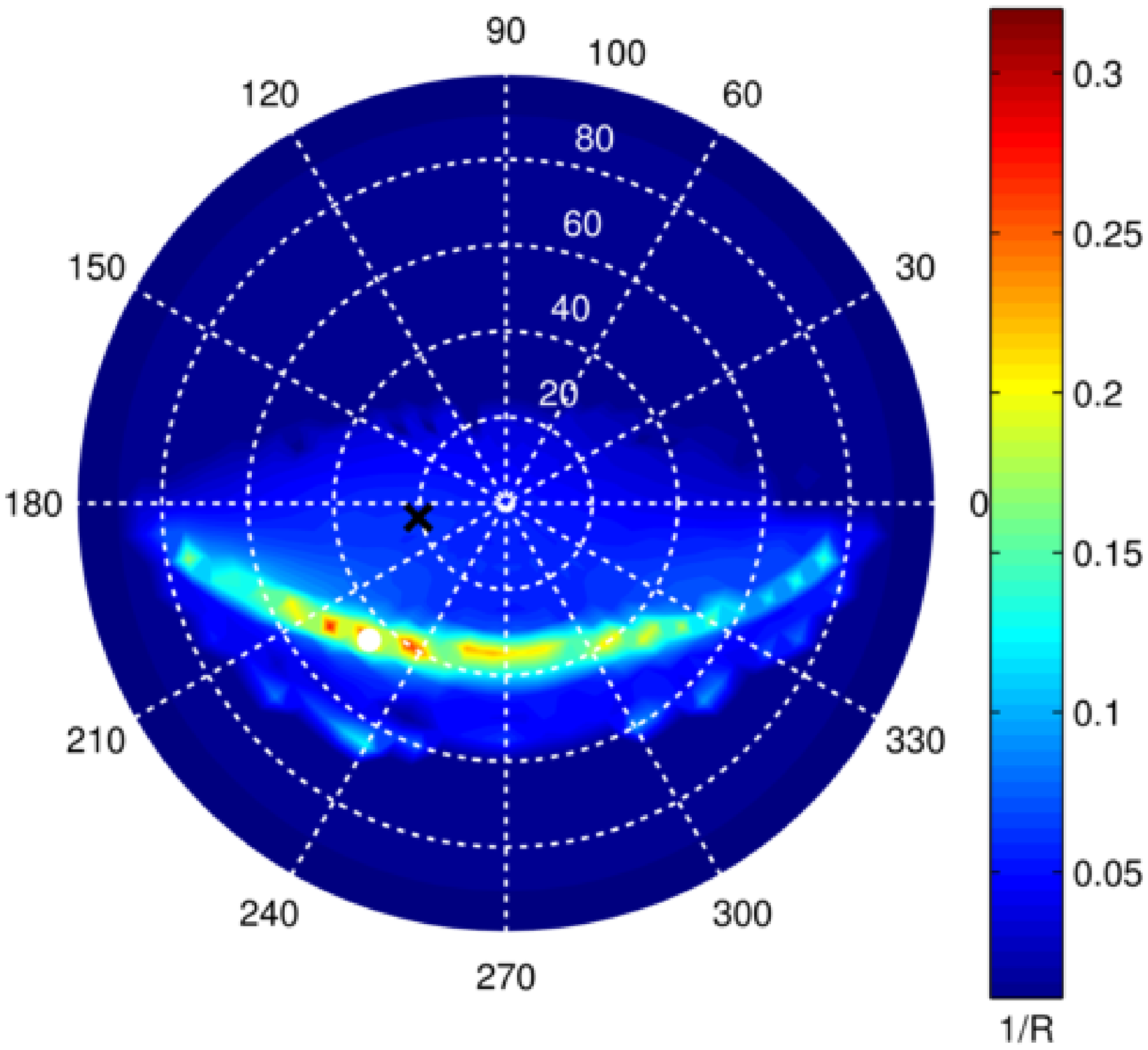}
\includegraphics[width=0.5\textwidth,clip=]{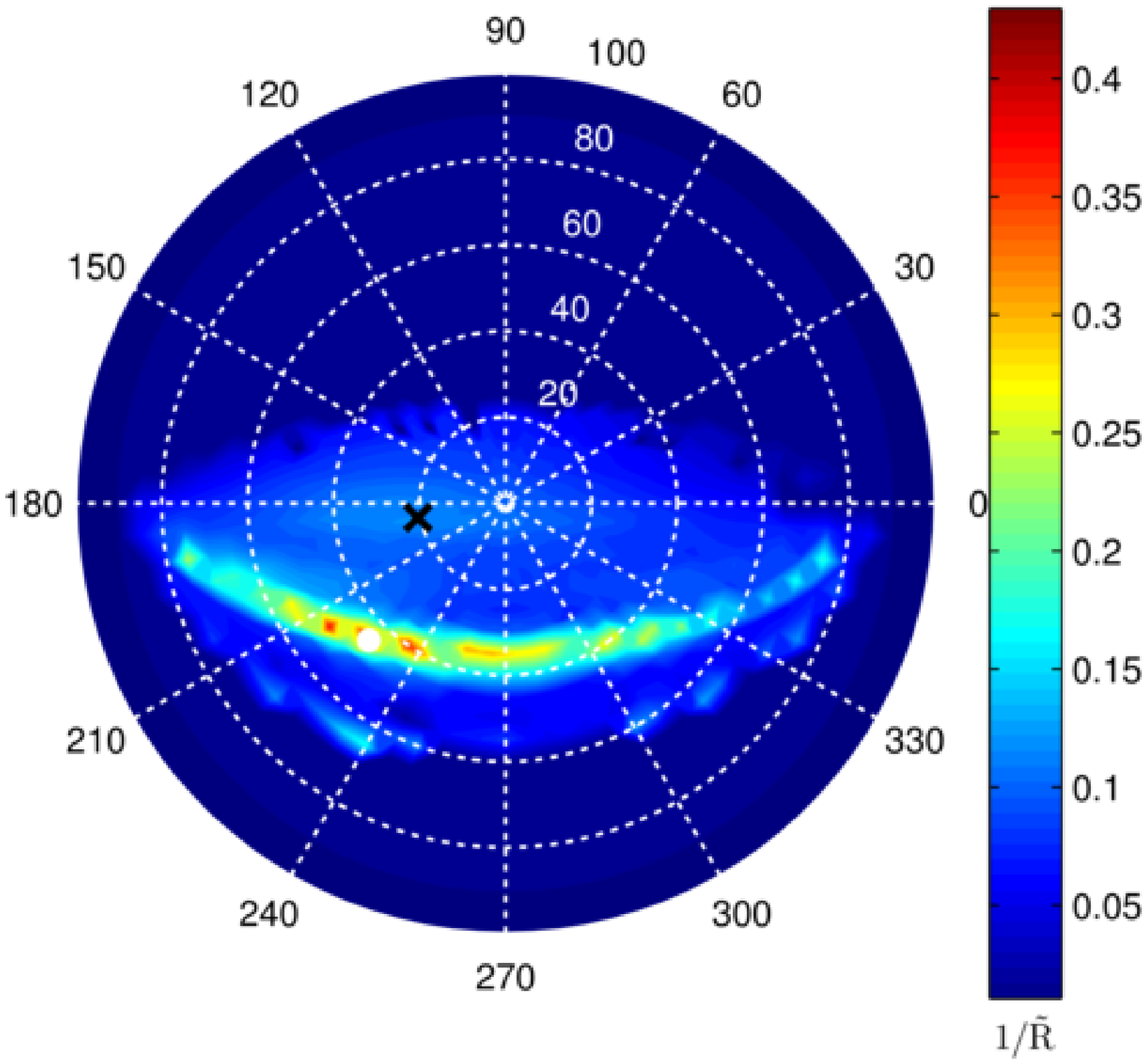}
}
\caption{Original (left) and filtered (right) residue maps for the 2009-07-11 STA event.}\label{fig:STA_2009-07-11_RM}
\end{figure}

\begin{figure}
\centerline{
\includegraphics[width=0.5\textwidth,clip=]{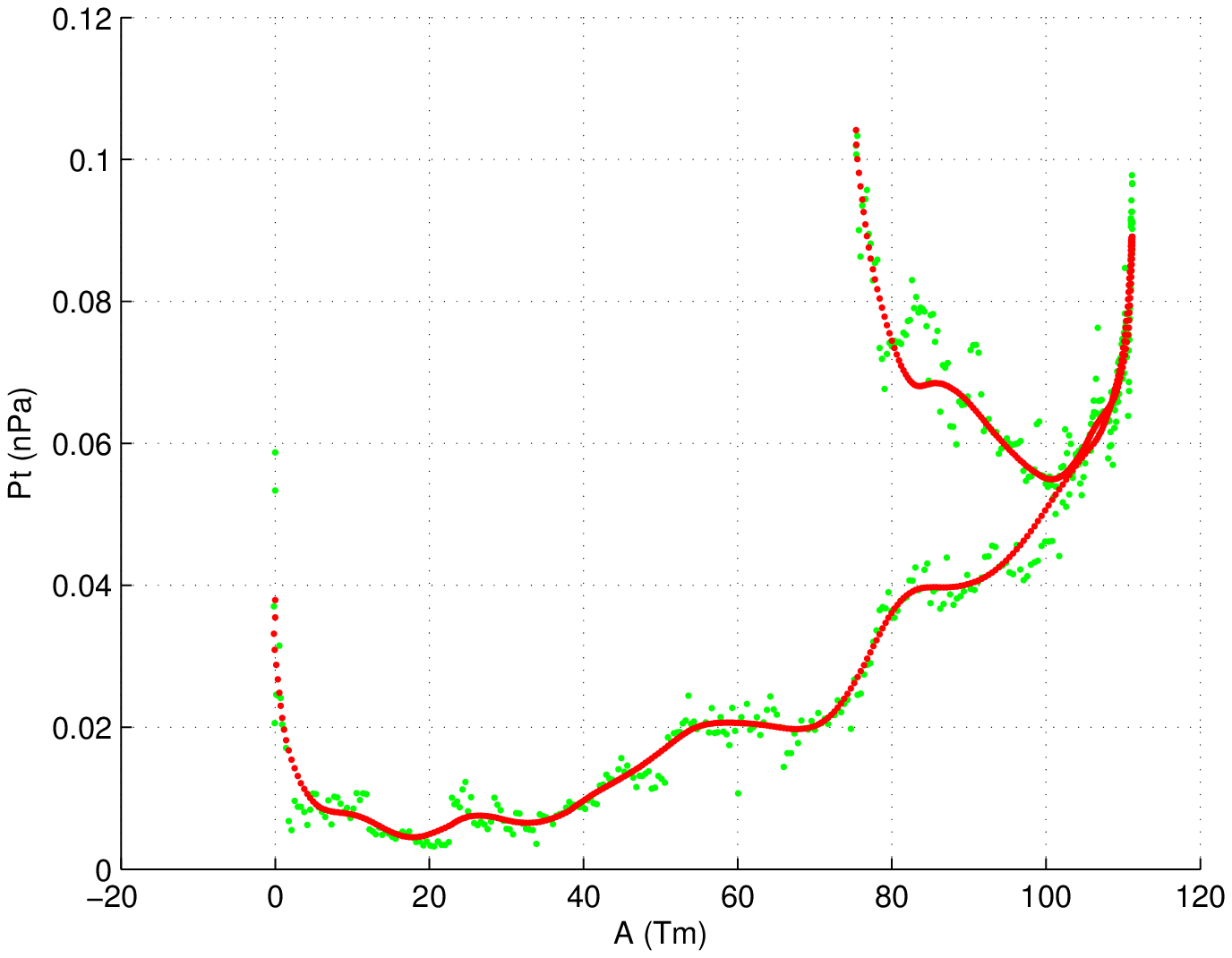}
\includegraphics[width=0.5\textwidth,clip=]{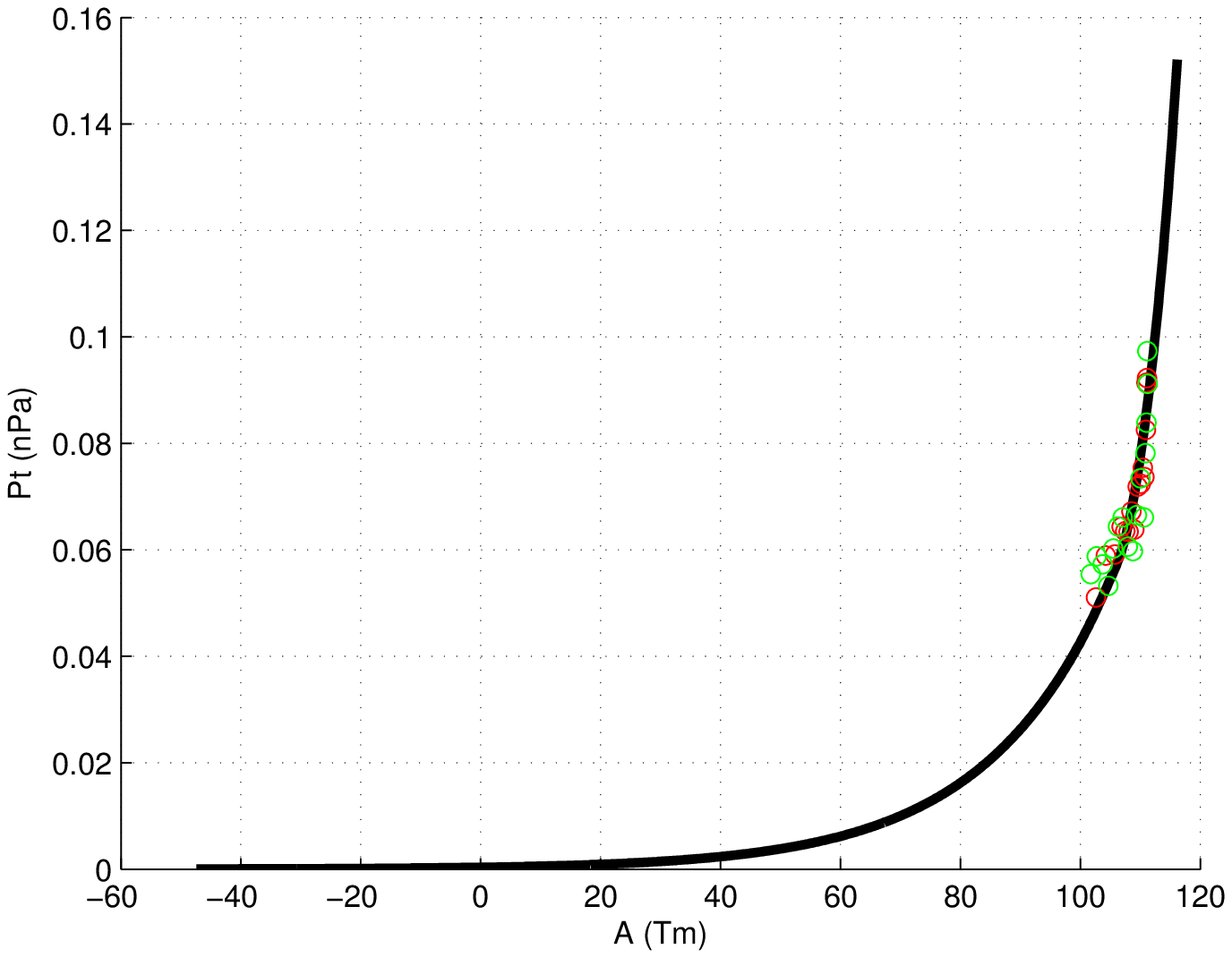}
}
\caption{The $P_t(A)$ curve smoothed with running average (left) and fitted with polynomial of the 2nd order and exponential tail (right) for the 2009-07-11 STA event.}\label{fig:STA_2009-07-11_AP}
\end{figure}

\begin{figure}
\centerline{
\includegraphics[width=0.5\textwidth,clip=]{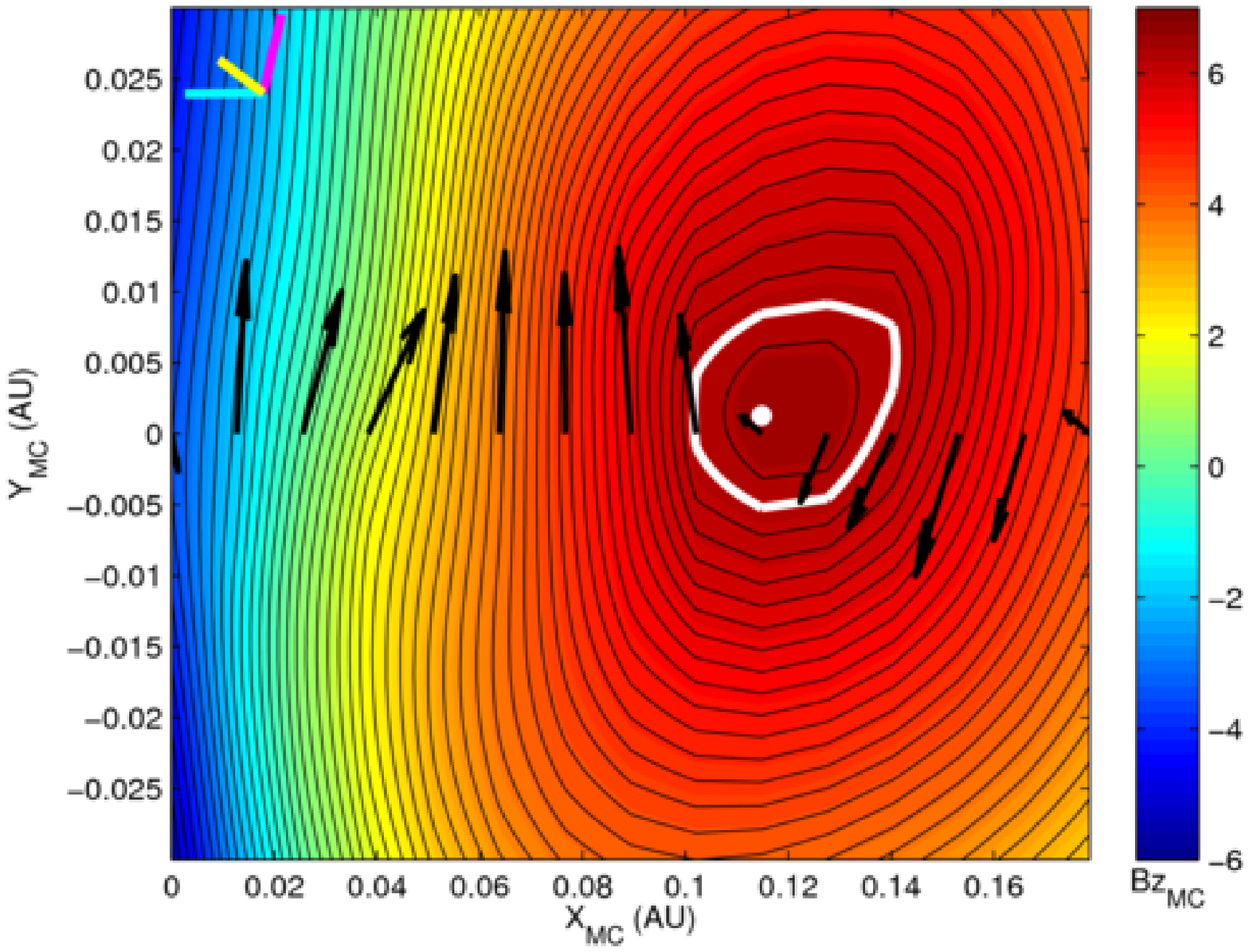}
}
\caption{The reconstructed magnetic field map for the 2009-07-11 STA event. The projected coordinates shown in the magnetic field map are $R$ (cyan), $T$ (magenta), $N$ (yellow).}\label{fig:STA_2009-07-11_GS}
\end{figure}

The MC observed on July 11th 2009 by the STEREO-A spacecraft is an example of an ICME followed by a stream interaction region. As seen in Figure~\ref{fig:STA_2009-07-11_data} the plasma pressure has a well distinguished maximum in the middle of the spacecraft transit through the MC, which indicates a small impact parameter for this event (Group 1 type of events according to \inlinecite{Jian2006}). Closer to the end of the time interval of a smooth magnetic field rotation the plasma pressure grows and well exceeds its value in the central part of the flux rope. This increase of plasma pressure is caused by fast solar wind pushing the MC from behind. This ICME originated from a CME event on July 7th 2009. The calculated speed of the deHoffmann-Teller frame was \mbox{$V_{HT} = [304.8; 4.5; -3.9]$~km/s} in the RTN coordinates with the correlation coefficient $c=0.999$. In such cases the algorithm for invariant axis estimation produces a $P_t(A)$ curve similar to that shown in Figure~\ref{fig:STA_2009-07-11_AP} (left). The inward and outward branches of $P_t(A)$ perfectly coincide in the central part but differ greatly in transverse pressure in the boundary of the flux rope. This happens because the fast solar wind causes transverse pressure increase in the rear part of the MC, distorting that part, while the front part of the MC remains undistorted. The GS reconstruction may be reliable only for the central part of the flux rope where both inward and outward parts of the magnetic field lines remain uninfluenced by the fast solar wind. This small part of the MC is indicated in Figure~\ref{fig:STA_2009-07-11_GS} as the boundary of the flux rope. The estimated orientation of the flux rope is \mbox{$\theta_{RTN}=-51.7^{\circ}$} and \mbox{$\varphi_{RTN}=33.1^{\circ}$} in RTN. The flux rope was crossed by the spacecraft at a rather sharp angle far from the apex. The impact parameter of the intersection is $0.001$~AU. For this event the invariant axis estimations produced by GSR and MVA techniques differ by $\sim 30^{\circ}$.

\section{Discussion}\label{s:Discussion} 
We have presented an overview of the Grad-Shafranov reconstruction technique and two improvements to the algorithm. The improvements are filtering of the residual map with the branch length and the estimation of the second order derivative of the magnetic potential $A$ used throughout the algorithm with a robust noise-free filter. We have also conducted a GSR analysis of three sample ICMEs that have been chosen to differ greatly in their properties: a clear well-defined MC, a fast MC and an MC followed by the fast solar wind. From the results of the analysis we see that the GSR gives relevant information about a particular MC, although like any data analysis method it should be applied with its limitations taken into account.

The magnetic field maps reconstructed for all three events are almost circular in shape. This contradicts the MHD modelling according to which a typical flux rope is pancake-shaped \cite{Riley2004kt}. This difference was pointed out by \inlinecite{Riley2004}, who conducted a blind test comparison of a simulated ICME with various flux rope models. One of the reasons for this apparent flaw in GSR may be in its assumptions, i.e. the magnetohydrostatic equilibrium and the time stationarity of the MC. The ill-defined mathematical problem solved in GSR may also be a reason. Since the Cauchy problem for the elliptic partial differential equation (\ref{eq:GS}) is solved numerically without boundary conditions, it is possible to obtain  almost any result for a given distance from the spacecraft trajectory by changing the size of the step along the $y$ direction. In \inlinecite{Hau1999} this step size was chosen to be ${\Delta y}/{\Delta x}=0.1$ based on benchmark tests. In the benchmarking an analytical solution of the equation $\nabla^{2}A=e^{-2A}$ was used. In our benchmark runs we have obtained better results with ${\Delta y}/{\Delta x}=0.05$. Determination of the $y$-step can only be made empirically for a particular reconstruction.

One possible way of estimating the boundaries of the reconstruction, and thus the $y$-step, is to fix the aspect ratio of the cross-section of the flux rope. For instance, in elliptical models the aspect ratio of 4-to-1 is often used to mimic the pancake shape of the flux rope cross-section ({\eg} Mulligan and Russel, \citeyear{Mulligan2001}). The other possible way of estimating the aspect ratio is based on coronagraph observations of CMEs. Since the flow of plasma in an MC perpendicular to the direction of its propagation is very slow (Owens \etal, \citeyear{Owens2006}) the angular size of the ICME remains constant. \inlinecite{Owens2008} suggested that coronagraph measuremants can be used for estimating the angular size of the CME, i.e. the CME width, close to the Sun. Knowing the CME width it is possible to estimate the aspect ratio of the flux rope cross-section and use it later as a boundary constraint for the GSR.

\begin{figure}
\centerline{
\includegraphics[width=0.5\textwidth,clip=]{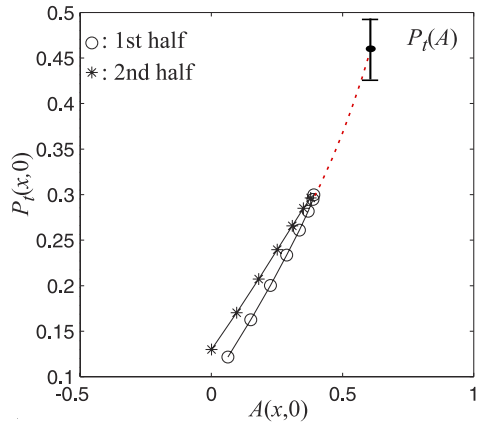}
}
\caption{A sketch showing uncertainties in $P_t(A)$ fitting for large impact parameters.}\label{fig:Pt_A_uncertainties}
\end{figure}

Most of the flux rope models are able to give reasonable results only for small impact parameters. The GSR possesses this disadvantage too. The reason for this is the uncertainty of the definition of the fitting curve for $P_t(A)$ (Figure~\ref{fig:Pt_A_uncertainties}). For large impact parameters the high transverse pressure part of the $P_t(A)$ curve is undefined for initial observational data and needs to be extrapolated. For various extrapolation functions the difference between values of transverse pressure for the same values of magnetic potential can well exceed orders of magnitude.

One more reason why the results of GSR may be questioned is that this method can show existence of a flux rope in an ICME even when there is none, which was pointed out by \inlinecite{Hasegawa2007}.

Despite these shortcomings GSR proves itself as a useful tool for estimating the critical parameters of MCs. In this paper we have made GS reconstruction for three sample ICMEs under very diverse solar wind conditions. In all these events the results of GSR agree with Jian's {\etal} (\citeyear{Jian2006}) classification of MCs. We were able to determine the invariant axis directions for these MCs. The axis directions were also estimated using minimum variance analysis (MVA). Note, that for the first event the direction of the invariant axis is close to the direction obtained in GSR (Figure \ref{fig:STA_2008-11-07_RM}), while for the second and the third events the invariant axis estimates given by GSR and MVA differ greatly (Figures \ref{fig:ACE_WIND_2004-11-09_RM_AP} (top and middle), \ref{fig:STA_2009-07-11_RM}). Thus, for well-defined MCs propagating in slow solar wind with a speed lower than the average speed of solar wind GSR and MVA techniques produce similar results for the invariant axis direction. While for the non-trivial cases of fast MCs and MCs distorted by high speed streams of solar wind GSR is capable of giving more relevant results.

%

%

%
\begin{acks}
The work of A. Isavnin and E. Kilpua was supported by the Academy of Finland.
\end{acks}

%
%
\bibliographystyle{spr-mp-sola}
\bibliography{bibliography.bib}  

\begin{thebibliography}{22}
\ifx \bisbn   \undefined \def \bisbn  #1{ISBN #1}\fi
\ifx \binits  \undefined \def \binits#1{#1}\fi
\ifx \bauthor  \undefined \def \bauthor#1{#1}\fi
\ifx \batitle  \undefined \def \batitle#1{#1}\fi
\ifx \bjtitle  \undefined \def \bjtitle#1{\textit{#1}}\fi
\ifx \bvolume  \undefined \def \bvolume#1{\textbf{#1}}\fi
\ifx \byear  \undefined \def \byear#1{#1}\fi
\ifx \bissue  \undefined \def \bissue#1{#1}\fi
\ifx \bfpage  \undefined \def \bfpage#1{#1}\fi
\ifx \blpage  \undefined \def \blpage #1{#1}\fi
\ifx \burl  \undefined \def \burl#1{\textsf{#1}}\fi
\ifx \href  \undefined \def \href#1#2{\textsf{#2}}\fi
\ifx \doiurl  \undefined \def
  \doiurl#1{\href{http://dx.doi.org/#1}{\textsf{#1}}}\fi
\ifx \betal  \undefined \def \betal{\textit{et al.}}\fi
\ifx \binstitute  \undefined \def \binstitute#1{#1}\fi
\ifx \bctitle  \undefined \def \bctitle#1{#1}\fi
\ifx \beditor  \undefined \def \beditor#1{#1}\fi
\ifx \bpublisher  \undefined \def \bpublisher#1{#1}\fi
\ifx \bbtitle  \undefined \def \bbtitle#1{\textit{#1}}\fi
\ifx \bedition  \undefined \def \bedition#1{#1}\fi
\ifx \bseriesno  \undefined \def \bseriesno#1{\textbf{#1}}\fi
\ifx \blocation  \undefined \def \blocation#1{#1}\fi
\ifx \bsertitle  \undefined \def \bsertitle#1{\textit{#1}}\fi
\ifx \bsnm \undefined \def \bsnm#1{#1}\fi
\ifx \bsuffix \undefined \def \bsuffix#1{#1}\fi
\ifx \bparticle \undefined \def \bparticle#1{#1}\fi
\ifx \barticle \undefined \def \barticle#1{}\fi
\ifx \botherref \undefined \def \botherref#1{}\fi
\ifx \url \undefined \def \url#1{\textsf{#1}}\fi
\ifx \bchapter \undefined \def \bchapter#1{}\fi
\ifx \bbook \undefined \def \bbook#1{}\fi
\ifx \bcomment \undefined \def \bcomment#1{#1}\fi
\ifx \oauthor \undefined \def \oauthor#1{#1}\fi
\ifx \citeauthoryear \undefined \def \citeauthoryear#1{#1}\fi
\def \endbibitem {}

\bibitem[\protect\citeauthoryear{{Burlaga}}{1988}]{Burlaga1988}
\begin{barticle}
\bauthor{\bsnm{{Burlaga}}, \binits{L.F.}}:
\byear{1988},
\batitle{{Magnetic clouds and force-free fields with constant alpha}}.
\bjtitle{\jgr}
\bvolume{93}(\bissue{A7}),
\bfpage{7217}\,--\,\blpage{7224}.
doi:\doiurl{10.1029/JA093iA07p07217}.
\end{barticle}
\endbibitem

\bibitem[\protect\citeauthoryear{{Burlaga} \textit{et~al.}}{1981}]{Burlaga1981}
\begin{barticle}
\bauthor{\bsnm{{Burlaga}}, \binits{L.}}, \bauthor{\bsnm{{Sittler}},
  \binits{E.}}, \bauthor{\bsnm{{Mariani}}, \binits{F.}},
  \bauthor{\bsnm{{Schwenn}}, \binits{R.}}:
\byear{1981},
\batitle{{Magnetic loop behind an interplanetary shock: Voyager, Helios and IMP
  8 observations}}.
\bjtitle{\jgr}
\bvolume{86}(\bissue{A8}),
\bfpage{6673}\,--\,\blpage{6684}.
\end{barticle}
\endbibitem

\bibitem[\protect\citeauthoryear{{Farrugia}
  \textit{et~al.}}{1993}]{Farrugia1993}
\begin{barticle}
\bauthor{\bsnm{{Farrugia}}, \binits{C.J.}}, \bauthor{\bsnm{{Richardson}},
  \binits{I.G.}}, \bauthor{\bsnm{{Burlaga}}, \binits{L.F.}},
  \bauthor{\bsnm{{Lepping}}, \binits{R.P.}}, \bauthor{\bsnm{{Osherovich}},
  \binits{V.A.}}:
\byear{1993},
\batitle{{Simultaneous observations of solar MeV particles in a magnetic cloud
  and in the Earth's northern tail lobe: Implications for the global field line
  topology of magnetic clouds and for the entry of solar particles into the
  magnetosphere during cloud passage}}.
\bjtitle{\jgr}
\bvolume{98}(\bissue{A9}),
\bfpage{15497}\,--\,\blpage{15507}.
doi:\doiurl{10.1029/93JA01462}.
\end{barticle}
\endbibitem

\bibitem[\protect\citeauthoryear{{Gosling} \textit{et~al.}}{1987}]{Gosling1987}
\begin{barticle}
\bauthor{\bsnm{{Gosling}}, \binits{J.T.}}, \bauthor{\bsnm{{Baker}},
  \binits{D.N.}}, \bauthor{\bsnm{{Bame}}, \binits{S.J.}},
  \bauthor{\bsnm{{Feldman}}, \binits{W.C.}}, \bauthor{\bsnm{{Zwickl}},
  \binits{R.D.}}, \bauthor{\bsnm{{Smith}}, \binits{E.J.}}:
\byear{1987},
\batitle{{Bidirectional Solar Wind Electron Heat Flux Events}}.
\bjtitle{\grl}
\bvolume{92}(\bissue{A8}),
\bfpage{8519}\,--\,\blpage{8535}.
doi:\doiurl{10.1029/JA092iA08p08519}.
\end{barticle}
\endbibitem

\bibitem[\protect\citeauthoryear{{Hasegawa}
  \textit{et~al.}}{2007}]{Hasegawa2007}
\begin{barticle}
\bauthor{\bsnm{{Hasegawa}}, \binits{H.}}, \bauthor{\bsnm{{Nakamura}},
  \binits{R.}}, \bauthor{\bsnm{{Fujimoto}}, \binits{M.}},
  \bauthor{\bsnm{{Sergeev}}, \binits{V.A.}}, \bauthor{\bsnm{{Lucek}},
  \binits{E.A.}}, \bauthor{\bsnm{{R{\'e}me}}, \binits{H.}},
  \bauthor{\bsnm{{Khotyaintsev}}, \binits{Y.}}:
\byear{2007},
\batitle{{Reconstruction of a bipolar magnetic signature in an earthward jet in
  the tail: Flux rope or 3D guide-field reconnection?}}
\bjtitle{\jgr}
\bvolume{112}.
doi:\doiurl{10.1029/2007JA012492}.
\end{barticle}
\endbibitem

\bibitem[\protect\citeauthoryear{{Hau} and {Sonnerup}}{1999}]{Hau1999}
\begin{barticle}
\bauthor{\bsnm{{Hau}}, \binits{L.N.}}, \bauthor{\bsnm{{Sonnerup}},
  \binits{B.U.{\"O}.}}:
\byear{1999},
\batitle{{Two-dimensional coherent structures in the magnetopause: Recovery of
  static equilibria from single-spacecraft data}}.
\bjtitle{\jgr}
\bvolume{104}(\bissue{A4}),
\bfpage{6899}\,--\,\blpage{6917}.
doi:\doiurl{10.1029/1999JA900002}.
\end{barticle}
\endbibitem

\bibitem[\protect\citeauthoryear{{Hidalgo}, {Nieves-Chinchilla}, and
  {Cid}}{2002}]{Hidalgo2002}
\begin{barticle}
\bauthor{\bsnm{{Hidalgo}}, \binits{M.A.}}, \bauthor{\bsnm{{Nieves-Chinchilla}},
  \binits{T.}}, \bauthor{\bsnm{{Cid}}, \binits{C.}}:
\byear{2002},
\batitle{{Elliptical cross-section model for the magnetic topology of magnetic
  clouds}}.
\bjtitle{\grl}
\bvolume{29},
\bfpage{1637}\,--\,\blpage{1640}.
doi:\doiurl{10.1029/2001GL013875}.
\end{barticle}
\endbibitem

\bibitem[\protect\citeauthoryear{{Holoborodko}}{2008}]{Holoborodko2008}
\begin{botherref}
\oauthor{\bsnm{{Holoborodko}}, \binits{P.}}:
2008,
{Smooth noise-robust differentiators}.
\url{http://www.holoborodko.com/pavel/?page_id=245}.
\end{botherref}
\endbibitem

\bibitem[\protect\citeauthoryear{{Hu} and {Sonnerup}}{2002}]{Hu2002}
\begin{barticle}
\bauthor{\bsnm{{Hu}}, \binits{Q.}}, \bauthor{\bsnm{{Sonnerup}},
  \binits{B.U.{\"O}.}}:
\byear{2002},
\batitle{{Reconstruction of magnetic clouds in the solar wind: Orientations and
  configurations}}.
\bjtitle{\jgr}
\bvolume{107},
\bfpage{1142}.
doi:\doiurl{10.1029/2001JA000293}.
\end{barticle}
\endbibitem

\bibitem[\protect\citeauthoryear{{Huttunen}, {Koskinen}, and
  {Schwenn}}{2002}]{Huttunen2002}
\begin{barticle}
\bauthor{\bsnm{{Huttunen}}, \binits{K.E.J.}}, \bauthor{\bsnm{{Koskinen}},
  \binits{H.E.J.}}, \bauthor{\bsnm{{Schwenn}}, \binits{R.}}:
\byear{2002},
\batitle{{Variability of magnetospheric storms driven by different solar wind
  perturbations}}.
\bjtitle{\jgr}
\bvolume{107}(\bissue{A7}),
\bfpage{1121}\,--\,\blpage{1128}.
doi:\doiurl{10.1029/2001JA900171}.
\end{barticle}
\endbibitem

\bibitem[\protect\citeauthoryear{{Jian} \textit{et~al.}}{2006}]{Jian2006}
\begin{barticle}
\bauthor{\bsnm{{Jian}}, \binits{L.}}, \bauthor{\bsnm{{Russel}}, \binits{C.T.}},
  \bauthor{\bsnm{{Luhmann}}, \binits{J.G.}}, \bauthor{\bsnm{{Skoug}},
  \binits{R.M.}}:
\byear{2006},
\batitle{{Properties of interplanetary coronal mass ejections at one AU during
  1995-2004}}.
\bjtitle{\solphys}
\bvolume{239},
\bfpage{393}\,--\,\blpage{436}.
doi:\doiurl{10.1007/s11207-006-0133-2}.
\end{barticle}
\endbibitem

\bibitem[\protect\citeauthoryear{{Khrabrov} and
  {Sonnerup}}{1998}]{Khrabrov1998}
\begin{bbook}
\bauthor{\bsnm{{Khrabrov}}, \binits{A.V.}}, \bauthor{\bsnm{{Sonnerup}},
  \binits{B.U.{\"O}.}}:
\byear{1998},
\bbtitle{{Analysis Methods for Multi-Spacecraft Data}},
\bpublisher{ESA Pub. Div.},
\blocation{Keplerlaan 1, 2200 AG Noordwijk, The Netherlands},
\bfpage{221}\,--\,\blpage{248}.
\bcomment{Chap. 9}.
\end{bbook}
\endbibitem

\bibitem[\protect\citeauthoryear{{Lepping}, {Jones}, and
  {Burlaga}}{1990}]{Lepping1990}
\begin{barticle}
\bauthor{\bsnm{{Lepping}}, \binits{R.P.}}, \bauthor{\bsnm{{Jones}},
  \binits{J.A.}}, \bauthor{\bsnm{{Burlaga}}, \binits{L.F.}}:
\byear{1990},
\batitle{{Magnetic field structure of interplanetary magnetic clouds at 1 AU}}.
\bjtitle{\jgr}
\bvolume{95}(\bissue{A8}),
\bfpage{11957}\,--\,\blpage{11965}.
doi:\doiurl{10.1029/JA095iA08p11957}.
\end{barticle}
\endbibitem

\bibitem[\protect\citeauthoryear{{Marubashi} and
  {Lepping}}{2007}]{Marubashi2007}
\begin{barticle}
\bauthor{\bsnm{{Marubashi}}, \binits{K.}}, \bauthor{\bsnm{{Lepping}},
  \binits{R.P.}}:
\byear{2007},
\batitle{{Long-duration magnetic clouds: a comparison of analyses using torus-
  and cylinder-shaped flux rope models}}.
\bjtitle{\annG}
\bvolume{25},
\bfpage{2453}\,--\,\blpage{2477}.
doi:\doiurl{10.5194/angeo-25-2453-2007}.
\end{barticle}
\endbibitem

\bibitem[\protect\citeauthoryear{{M{\"o}stl}
  \textit{et~al.}}{2008}]{Moestl2008}
\begin{barticle}
\bauthor{\bsnm{{M{\"o}stl}}, \binits{C.}}, \bauthor{\bsnm{{Miklenic}},
  \binits{C.}}, \bauthor{\bsnm{{Farrugia}}, \binits{C.J.}},
  \bauthor{\bsnm{{Temmer}}, \binits{M.}}, \bauthor{\bsnm{{Veronig}},
  \binits{A.}}, \bauthor{\bsnm{{Galvin}}, \binits{A.B.}},
  \bauthor{\bsnm{{Vr{\v{s}}nak}}, \binits{B.}}, \bauthor{\bsnm{{Biernat}},
  \binits{H.K.}}:
\byear{2008},
\batitle{{Two-spacecraft reconstruction of a magnetic cloud and comparison to
  its solar source}}.
\bjtitle{\annG}
\bvolume{26},
\bfpage{3139}\,--\,\blpage{3152}.
doi:\doiurl{10.5194/angeo-26-3139-2008}.
\end{barticle}
\endbibitem

\bibitem[\protect\citeauthoryear{{Mulligan} and {Russel}}{2001}]{Mulligan2001}
\begin{barticle}
\bauthor{\bsnm{{Mulligan}}, \binits{T.}}, \bauthor{\bsnm{{Russel}},
  \binits{C.T.}}:
\byear{2001},
\batitle{{Multispacecraft modeling of the flux rope strucure of interplanetary
  coronal mass ejections: cylindrically symmetric versus nonsymmetric
  topologies}}.
\bjtitle{\jgr}
\bvolume{106}(\bissue{A6}),
\bfpage{10581}\,--\,\blpage{10596}.
doi:\doiurl{10.1029/2000JA900170}.
\end{barticle}
\endbibitem

\bibitem[\protect\citeauthoryear{{Owens}}{2008}]{Owens2008}
\begin{barticle}
\bauthor{\bsnm{{Owens}}, \binits{M.J.}}:
\byear{2008},
\batitle{{Combining remote and in-situ observations of coronal mass ejections
  to better constrain magnetic cloud reconstruction}}.
\bjtitle{\jgr}
\bvolume{113}(\bissue{12102}).
doi:\doiurl{10.1029/2008JA013589}.
\end{barticle}
\endbibitem

\bibitem[\protect\citeauthoryear{{Owens}, {Merkin}, and
  {Riley}}{2006}]{Owens2006}
\begin{barticle}
\bauthor{\bsnm{{Owens}}, \binits{M.J.}}, \bauthor{\bsnm{{Merkin}},
  \binits{V.G.}}, \bauthor{\bsnm{{Riley}}, \binits{P.}}:
\byear{2006},
\batitle{{A kinematically distorted flux rope model for magnetic clouds}}.
\bjtitle{\jgr}
\bvolume{111}(\bissue{A03104}).
doi:\doiurl{10.1029/2005JA011460}.
\end{barticle}
\endbibitem

\bibitem[\protect\citeauthoryear{{Richardson} and
  {Cane}}{2010}]{Richardson2010}
\begin{barticle}
\bauthor{\bsnm{{Richardson}}, \binits{I.G.}}, \bauthor{\bsnm{{Cane}},
  \binits{H.V.}}:
\byear{2010},
\batitle{{Near-Earth interplanetary coronal mass ejections during solar cycle
  23 (1996-2009): catalog and summary of properties}}.
\bjtitle{\solphys}
\bvolume{264},
\bfpage{189}\,--\,\blpage{237}.
doi:\doiurl{10.1007/s11207-010-9568-6}.
\end{barticle}
\endbibitem

\bibitem[\protect\citeauthoryear{{Riley} and {Crooker}}{2004}]{Riley2004kt}
\begin{barticle}
\bauthor{\bsnm{{Riley}}, \binits{P.}}, \bauthor{\bsnm{{Crooker}},
  \binits{N.U.}}:
\byear{2004},
\batitle{{Kinematic treatment of coronal mass ejection evolution in the solar
  wind}}.
\bjtitle{\apj}
\bvolume{600}(\bissue{2}),
\bfpage{1035}.
doi:\doiurl{10.1086/379974}.
\end{barticle}
\endbibitem

\bibitem[\protect\citeauthoryear{{Riley} \textit{et~al.}}{2004}]{Riley2004}
\begin{barticle}
\bauthor{\bsnm{{Riley}}, \binits{P.}}, \bauthor{\bsnm{{Linker}}, \binits{J.}},
  \bauthor{\bsnm{{Lionello}}, \binits{R.}}, \bauthor{\bsnm{{Miki{\'c}}},
  \binits{Z.}}, \bauthor{\bsnm{{Odstrcil}}, \binits{D.}},
  \bauthor{\bsnm{{Hidalgo}}, \binits{M.}}, \bauthor{\bsnm{{Cid}}, \binits{C.}},
  \bauthor{\bsnm{{Hu}}, \binits{Q.}}, \bauthor{\bsnm{{Lepping}}, \binits{R.}},
  \bauthor{\bsnm{{Lynch}}, \binits{B.}}, \bauthor{\bsnm{{Rees}}, \binits{A.}}:
\byear{2004},
\batitle{{Fitting flux ropes to a global MHD solution: a comparison of
  techniques}}.
\bjtitle{\jastp}
\bvolume{66},
\bfpage{1321}\,--\,\blpage{1331}.
doi:\doiurl{10.1016/j.jastp.2004.03.019}.
\end{barticle}
\endbibitem

\bibitem[\protect\citeauthoryear{{Zurbuchen} and
  {Richardson}}{2006}]{Zurbuchen2006}
\begin{barticle}
\bauthor{\bsnm{{Zurbuchen}}, \binits{T.H.}}, \bauthor{\bsnm{{Richardson}},
  \binits{I.G.}}:
\byear{2006},
\batitle{{In-situ solar wind and magnetic field signatures of interplanetary
  coronal mass ejections}}.
\bjtitle{\ssr}
\bvolume{123},
\bfpage{31}\,--\,\blpage{43}.
doi:\doiurl{10.1007/s11214-006-9010-4}.
\end{barticle}
\endbibitem

\end{thebibliography}
%
%
%
%

\end{article} 
\end{document}